\documentclass[reprint,superscriptaddress,amsmath,amssymb,aps,longbibliography]{revtex4-1}

\usepackage[colorlinks]{hyperref}
\usepackage{graphicx}
\usepackage{dcolumn}
\usepackage{bm}
\usepackage{booktabs}
\usepackage{array}
\usepackage{xcolor}
\usepackage{braket}
\usepackage{algorithm} 
\usepackage{algpseudocode} 
\usepackage{float}
\usepackage{braket}
\newcommand{\brakket}[2]{\langle #1 | #2 \rangle}

\def\avg#1{\mathinner{\langle{#1}\rangle}}
\newcommand{\eq}[1]{Eq.~\hyperref[eq:#1]{(\ref*{eq:#1})}}
\renewcommand{\sec}[1]{\hyperref[sec:#1]{Section~\ref*{sec:#1}}}
\newcommand{\app}[1]{\hyperref[app:#1]{Appendix~\ref*{app:#1}}}
\newcommand{\tab}[1]{\hyperref[tab:#1]{Table~\ref*{tab:#1}}}
\newcommand{\fig}[1]{\hyperref[fig:#1]{Fig.~\ref*{fig:#1}}}
\newcommand{\figs}[1]{\hyperref[fig:#1]{Fig. S\ref*{fig:#1}}}
\newcommand{\figa}[2]{\hyperref[fig:#1]{Fig.~\ref*{fig:#1}#2}}
\newcommand{\figx}[2]{\hyperref[fig:#1]{Fig.~\ref*{fig:#1}(#2)}}
\newcommand{\thm}[1]{\hyperref[thm:#1]{Theorem~\ref*{thm:#1}}}
\newcommand{\lem}[1]{\hyperref[lem:#1]{Lemma~\ref*{lem:#1}}}
\newcommand{\cor}[1]{\hyperref[cor:#1]{Corollary~\ref*{cor:#1}}}
\newcommand{\defn}[1]{\hyperref[def:#1]{Definition~\ref*{def:#1}}}
\newcommand{\alg}[1]{\hyperref[alg:#1]{Algorithm~\ref*{alg:#1}}}

\newcommand{\nick}[1]{\textcolor{black}{#1}}

\begin{document}

 \title{Hartree-Fock on a superconducting qubit quantum computer}
 
\author{Google AI Quantum and Collaborators}
\email[Corresponding author (Nicholas Rubin): ]{nickrubin@google.com}
\email[Corresponding author (Ryan Babbush): ]{babbush@google.com}

\date{\today}

\begin{abstract}
As the search continues for useful applications of noisy intermediate scale quantum devices, variational simulations of fermionic systems remain one of the most promising directions. Here, we perform a series of quantum simulations of chemistry which involve twice the number of qubits and more than ten times the number of gates as the largest prior experiments. We model the binding energy of ${\rm H}_6$, ${\rm H}_8$, ${\rm H}_{10}$ and ${\rm H}_{12}$ chains as well the isomerization of diazene. We also demonstrate error-mitigation strategies based on $N$-representability which dramatically improve the effective fidelity of our experiments. Our parameterized ansatz circuits realize the Givens rotation approach to noninteracting fermion evolution, which we variationally optimize to prepare the Hartree-Fock wavefunction. This ubiquitous algorithmic primitive corresponds to a rotation of the orbital basis and is required by many proposals for correlated simulations of molecules and Hubbard models. Because noninteracting fermion evolutions are classically tractable to simulate, yet still generate highly entangled states over the computational basis, we use these experiments to benchmark the performance of our hardware while establishing a foundation for scaling up more complex correlated quantum simulations of chemistry.
\end{abstract}

\maketitle

The prediction of molecular properties and chemical reactions from ab initio quantum mechanics has emerged as one of the most promising applications of quantum computing \cite{Aspuru-Guzik2005}. This \nick{fact} is due both to the commercial value of accurate simulations as well as the relatively modest number of qubits required to represent interesting instances. However, as the age of “quantum supremacy” dawns \cite{google_quantum_supremacy},
so has a more complete appreciation of the challenges required to scale such computations to the classically intractable regime using near-term intermediate scale quantum (NISQ) devices. Achieving that objective will require further algorithmic innovations, hardware with more qubits and low error rates, and more effective error-mitigation strategies. Here, we report a variational quantum eigensolver (VQE) \cite{Peruzzo2013} simulation of molecular systems with progress in all three directions.

\begin{figure}[t]
    \centering
    \includegraphics[width=8.5cm]{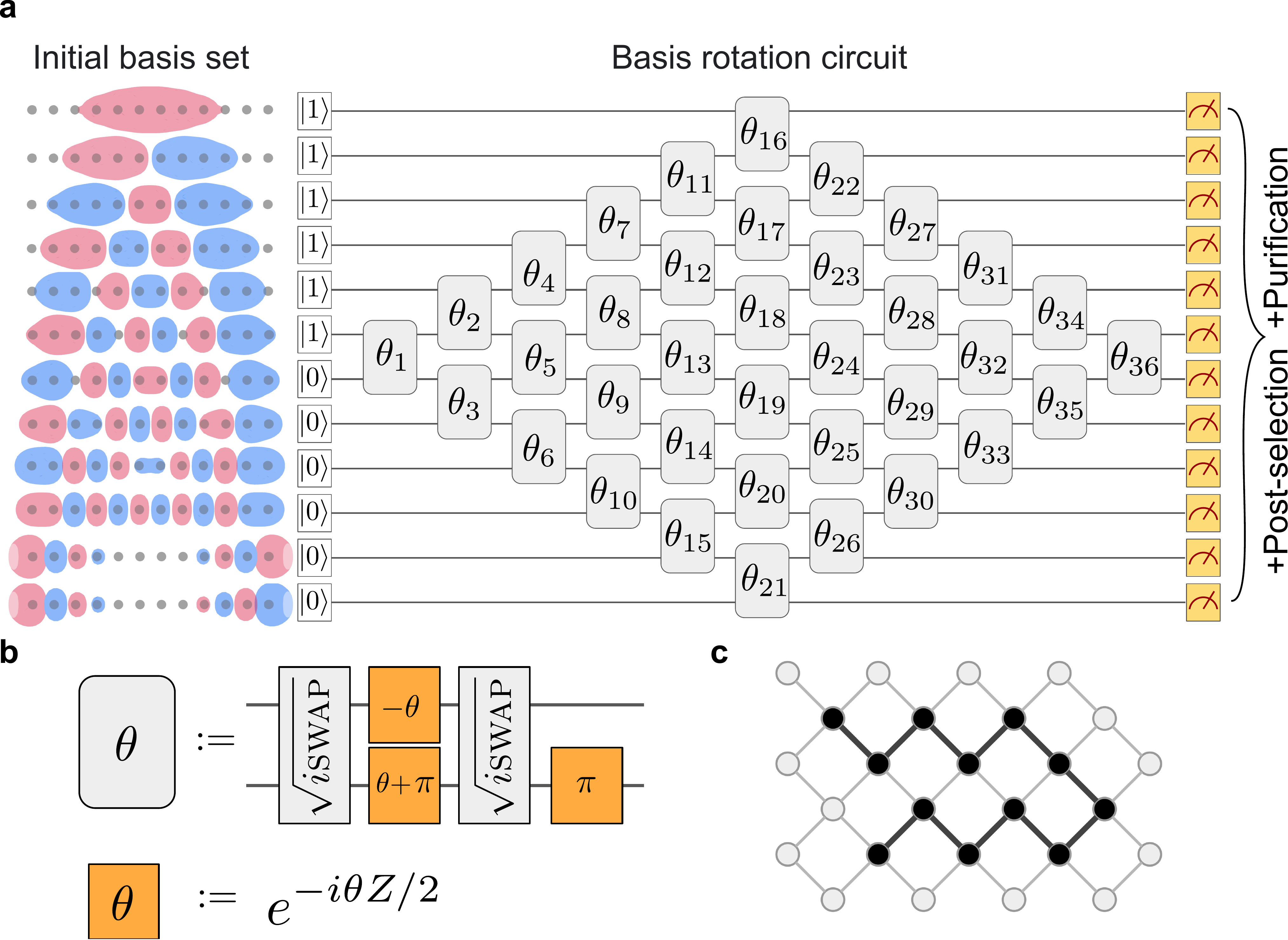}
\caption{\textbf{Basis rotation circuit and compilation.} a) To the left of the circuit diagram are the initial orbitals for the ${\rm H}_{12}$ chain with atom spacings of 1.3\! \AA, obtained by diagonalizing \nick{the} Hamiltonian ignoring electron-electron \nick{interactions}. The circuit diagram depicts the basis rotation ansatz for a linear chain of twelve hydrogen atoms. Each grey box with a rotation angle $\theta$ represents a Givens rotation gate.  b) Compilation of the Givens rotation gate to $\sqrt{i\textsc{swap}}$ gates and single-qubit gates that \nick{can be realized} directly in hardware. The $\rm{H}_{12}$ circuit involves $72$ $\sqrt{i\textsc{swap}}$ gates and $108$ single-qubit $Z$ rotation gates with a total of $36$ variational parameters. c) Depiction of a twelve qubit line on a subgrid of the entire $54$-qubit Sycamore device. All circuits only require gates between pairs of qubits which are adjacent in a linear topology.
    \label{fig:circuits_and_gates}}
\end{figure}

We use\nick{d} the Google Sycamore quantum processor to simulate the binding energy of hydrogen chains as large as ${\rm H}_{12}$, as well as a chemical reaction mechanism (the isomerization of diazene). The Sycamore quantum processor consists of a two-dimensional array of $54$ transmon  qubits~\cite{google_quantum_supremacy}.   Each  qubit is tunably coupled to four  nearest  neighbors  in  a  rectangular  lattice.   Our largest simulations used a dozen qubits -- twice the size as the largest prior quantum simulations of chemistry~\cite{kandala2017hardware} -- and require\nick{d} only nearest-neighbor coupling (depicted in \fig{circuits_and_gates}).  Prior simulations of chemistry on superconducting qubit devices and trapped ion systems demonstrated the possibility of error mitigation through VQE~\cite{OMalley2016,kandala2017hardware,PhysRevX.8.031022,kandala2019error,Siddiqi2017,PhysRevA.100.022517,PhysRevA.100.010302}, albeit on a small scale. We demonstrated  that, to within the model, achieving chemical accuracy through VQE is possible for intermediate scale problems when combined with effective error mitigation strategies.  Furthermore, we argue that the circuit ansatz we use\nick{d} for VQE is especially appealing as a benchmark for chemistry.

We simulate\nick{d} quantum chemistry in a second-quantized representation where the state of each of $N$ qubits encode\nick{d} the occupancy of an orbital basis function. We use\nick{d} what are commonly referred to as “core orbitals” as the initial orbitals (shown for ${\rm H}_{12}$ on the left of \figa{circuits_and_gates}{a}), which are the eigenfunctions of the molecular Hamiltonian without the electron-electron interaction term. The goal of this experiment \nick{was} to use a quantum computer to implement the Hartree-Fock procedure, which is a method for obtaining the best single-particle orbital functions assuming each electron feels the average potential generated from all the other electrons.  This assumption is enforced by constraining the wavefunction to be a product of one-particle functions which has been appropriately \nick{antisymmetrized} to satisfy the Pauli exclusion principle. An initial guess for the Hartree-Fock state, from which we can optimize the orbitals, was obtained by filling the lowest energy $\eta / 2$ orbitals, each with a spin-up electron and a spin-down electron, where $\eta$ is the number of electrons. Since we simulate\nick{d} the singlet ground state for all molecules considered here, there is no spin component to the mean-field approximation; thus, we only need\nick{ed} to explicitly simulate the $\eta/2$ spin-up electrons.

By performing a unitary rotation of the initial (core) orbital basis $\varphi_p(r)$, one can obtain a new valid set of orbitals $\widetilde{\varphi}_p(r)$ as a linear combination of the initial ones:
\begin{equation}
\label{eq:basis_change}
\widetilde{\varphi}_p\!\left(r\right) = \sum_{q=1}^N \left[e^\kappa\right]_{pq} \varphi_q\!\left(r\right),
\end{equation}
where $\kappa$ is an $N \times N$ anti-Hermitian matrix and $[e^\kappa]_{pq}$ is the $p,q$ element of the matrix exponential of $\kappa$. \nick{A result due to Thouless} \cite{Thouless1960} is that one can express the unitary that applies this basis rotation to the quantum state as time-evolution under a \nick{non-interacting} fermion Hamiltonian. Specifically, if we take $a^\dagger_p$ and $a_p$ to be fermionic creation and annihilation operators for the core orbital $\varphi_p(r)$ then we can parameterize $\ket{\psi_\kappa}$, an antisymmetric product state in the new basis $\widetilde{\varphi}_p(r)$, as \nick{non-interacting} fermion dynamics from a computational basis state $\ket{\eta} = a^\dagger_\eta \cdots a^\dagger_1 \ket{0}$ in the core orbital basis:
\begin{equation}
\label{eq:U}
\ket{\psi_\kappa} = U_\kappa \ket{\eta}, \qquad  U_\kappa = \exp\left(\sum_{p,q=1}^N \kappa_{pq} a^\dagger_p a_q\right).
\end{equation}
Such states are referred to as Slater determinants.

\nick{To complete the accurate preparation of Hartree-Fock states, we implemented variational relaxation of the $\kappa$ parameters to minimize the energy of $\ket{\psi_\kappa}$ starting from the optimal $\kappa$ determined by solving the Hartree-Fock equations classically.  This is an idealized implementation of VQE that allowed us to demonstrate error mitigation of coherent errors through variational relaxation.}
We define\nick{d} the Hartree-Fock state $\ket{\psi_{\rm HF}}$ to be the lowest energy Slater determinant for the molecular Hamiltonian $H$, i.e.
\begin{equation}
\label{eq:hartree-fock}
\ket{\psi_{\rm HF}} = \ket{\psi_{\kappa^\star}} 
\qquad \kappa^\star = \textrm{argmin}_\kappa
\bra{\psi_\kappa} H \ket{\psi_\kappa}.
\end{equation}
We \nick{applied} $U_\kappa$ to $\ket{\eta}$ using our quantum computer and then perform\nick{ed} the optimization over $\kappa$ through feedback from a classical optimization routine. The energy decrease\nick{d} because the initial core orbitals were obtained by ignoring the electron-electron interaction \nick{and variational relaxation compensates for coherent errors}. \nick{Since the generator for $U_\kappa$ corresponds to a \nick{non-interacting} fermion Hamiltonian, its action on a product state in second quantization can be classically simulated in $O(N^3)$ by diagonalizing the one-body operator and in some cases the Hartree-Fock procedure can be made to converge with even lower complexity}. Despite that fact, we argue that this \nick{procedure} is still a compelling experiment for a quantum computer.

\begin{figure*}[htb]
    \centering 
    \includegraphics[width=0.99\textwidth]{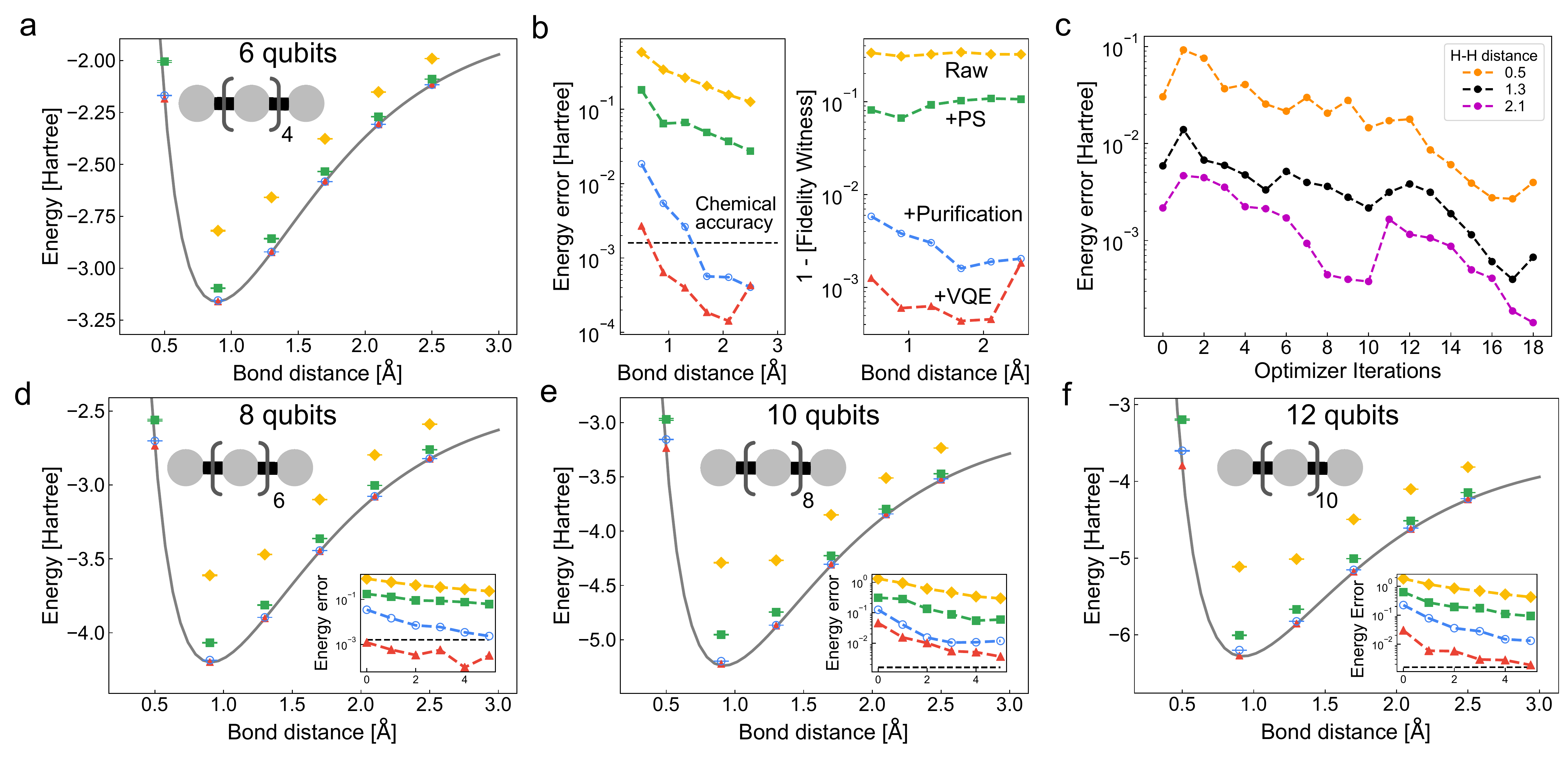}
    \caption{\textbf{Static and VQE performance on \nick{h}ydrogen chains.} Binding curve simulations for $\rm{H}_{6}$, $\rm{H}_{8}$, $\rm{H}_{10}$, and $\rm{H}_{12}$ with various forms of error mitigation.  Subfigures (a, d, e, f) compare Sycamore's \nick{raw performance (yellow diamonds)} with post-selection (green squares), purification (blue circles), and error mitigated combined with variational relaxation (red triangles). \nick{For all hydrogen systems the raw data at 0.5 \AA\, bond length is off the top of the plot.}  The yellow, green, and blue points were calculated using the optimal basis rotation angles computed from a classical simulation; thus, the variational optimization shown here is only used to correct systematic errors in the circuit realization.  Subfigure (b) contains the absolute error and infidelity for the $\rm{H}_{6}$ system. \nick{For all points we calculated} a fidelity witness described in \app{fidelity_fidelity_witness}.  \nick{The error bars for all points were computed by estimating the covariance between simultaneously measured sets of $1$-RDM elements and resampling those elements under a multivariate Gaussian model.  Energies from each sample were tabulated and the standard deviation is used as the error bar.} The ``+PS'' means applying post-selection to the raw data, ``+Purification'' means applying post-selection and McWeeny purification, and ``+VQE'' means post-selection, McWeeny purification, and variational relaxation.  Subfigure (c) contains optimization traces for three $\rm{H}_{6}$ geometries (bond distances of 0.5 \AA, 1.3 \AA, and 2.1 \AA). \nick{All optimization runs used between 18 and 30 iterations.  The lowest energy solution from the optimization trace was reported.}}
    \label{fig:fig2_h6_curve}
\end{figure*}

The Hartree-Fock state is usually the initial state for classical correlated electronic structure calculations 
such as coupled cluster and configuration interaction \nick{methods}, as well as for many quantum algorithms for chemistry. Thus, often one chooses to work in the molecular orbital basis, which is defined so that the Hartree-Fock state is a computational basis state. However, the molecular orbital basis Hamiltonian has a large number of terms which can be challenging to simulate and measure with low complexity. Accordingly, the most efficient quantum algorithms for chemistry \cite{Low2018,BabbushContinuum,childs2019theory,BabbushSpectra} require that one perform the simulation in more structured bases with asymptotically fewer terms \cite{BabbushLow,White2017,mcclean2019discontinuous}, necessitating that $U_{\kappa^\star}$ is applied explicitly at the beginning of the computation. Even when simulating chemistry in an arbitrary basis, the most efficient strategies are based on a tensor factorization of the Hamiltonian which requires many applications of $U_\kappa$ to simulate \cite{Motta2018,Berry2019}. Exploiting this tensor factorization with basis rotations is also key to the most efficient strategy for measuring $\avg{H}$ in variational algorithms, and requires implementing $U_\kappa$ prior to measurement \cite{huggins2019efficient}.

We use\nick{d} this variational ansatz based on basis rotations to benchmark the Sycamore processor for linear hydrogen chains of length $6$, $8$, $10$, and $12$ and two pathways for diazene bond isomerization. We model\nick{ed} hydrogen chains of length $N$ with $N$ qubits.  \nick{Our simulations required $N$ qubits to simulate $2N$ spin-orbitals due to the constraint that the $\alpha$-spin-orbitals have the same spatial wavefunction as the $\beta$-spin-orbitals.} For diazene we require\nick{d} $10$ qubits after pre-processing.  The hydrogen chains are a common benchmark in electronic structure~\cite{PhysRevX.7.031059, limacher2013new, hachmann2006multireference} and the diazene bond isomerization provides a system where the required accuracy is more representative of typical electronic structure problems and has been used as a benchmark for coupled cluster \nick{methods}~\cite{chaudhuri2008potential}.  For the diazene isomerization our goal was to resolve the energetic difference between the transition states of two competing mechanisms, requiring accuracy of about 40 milliHartree. This objective differs from prior quantum simulations of chemistry which have focused on bond dissociation curves\nick{~\cite{OMalley2016,kandala2017hardware,PhysRevX.8.031022,kandala2019error}}.

One motivation for this work \nick{was} to calibrate and validate the performance of our device in realizing an important algorithmic primitive for quantum chemistry and lattice model simulation. Our experiment \nick{was} also appealing for benchmarking purposes since the circuits we explore\nick{d} generate\nick{d} highly entangled states but with special structure that enable\nick{d} the efficient measurement of fidelity and the determination of systematic errors. Further motivation \nick{was} to implement the largest variational quantum simulation of chemistry so that \nick{it is possible to} better quantify the current gap between the capabilities of NISQ devices and real applications. Even though the Hartree-Fock ansatz is efficient to simulate classically, the circuits in our experiment are far more complex than prior experimental quantum simulations of chemistry. Finally, the structure of the Hartree-Fock state enable\nick{d} us to sample the energy and gradients of the variational ansatz with fewer measurements than would typically be required, allowing us to focus on other aspects of quantum simulating chemistry at scale, such as the effectiveness of various types of error-mitigation. Thus, our choice to focus on Hartree-Fock for this experiment embraces the notion that we should work towards valuable quantum simulations of chemistry by first scaling up important components of the exact solution (e.g., error-mitigation strategies and basis rotations) in a fashion that enables us to completely understand and perfect those primitives.

Variational algorithms are specified in the form of a functional minimization.  This minimization has three main components: ansatz specification in the form of a parameterized quantum circuit (the function), observable estimation (the functional), and outer-loop optimization (the minimization).  Each component is \nick{distinctively} affected by our choice to simulate a model corresponding to \nick{non-interacting} fermion wavefunctions.  Symmetries built into this ansatz allowed for reduction of the number of qubits required to simulate molecular systems, a reduction in the number of measurements needed to estimate the energy, and access to the gradient without additional measurements beyond those required for energy estimation. See \app{hf_via_ct} for details on how we realized Hartree-Fock with VQE. 

The unitary in \eq{U} can be compiled exactly (without Trotterization) using a procedure based on Givens rotations. This strategy was first suggested for quantum computing in work on linear optics in \cite{PhysRevLett.73.58} and later in the context of fermionic simulations in \cite{Wecker2015b}. Here, we implement\nick{ed} these basis rotations using the optimal compilation of \cite{Kivlichan2017} that has gate depth $N / 2$ and requires only $\eta (N - \eta)$ two qubit ``Givens rotation'' gates on a linearly connected architecture, giving one rotation for each element in the unitary basis change. These Givens rotation gates were implemented by decomposition into two $\sqrt{i\textsc{swap}}$ gates and three $\mathrm{Rz}$ gates. In \fig{circuits_and_gates}, we depict the basis change circuit for the $\textrm{H}_{12}$ chain, which has a diamond shaped structure. We further review the compilation of these circuits in \app{optimal_compilation}. 

\begin{figure}[tb]
    \centering 
    \includegraphics[width=8.5cm]{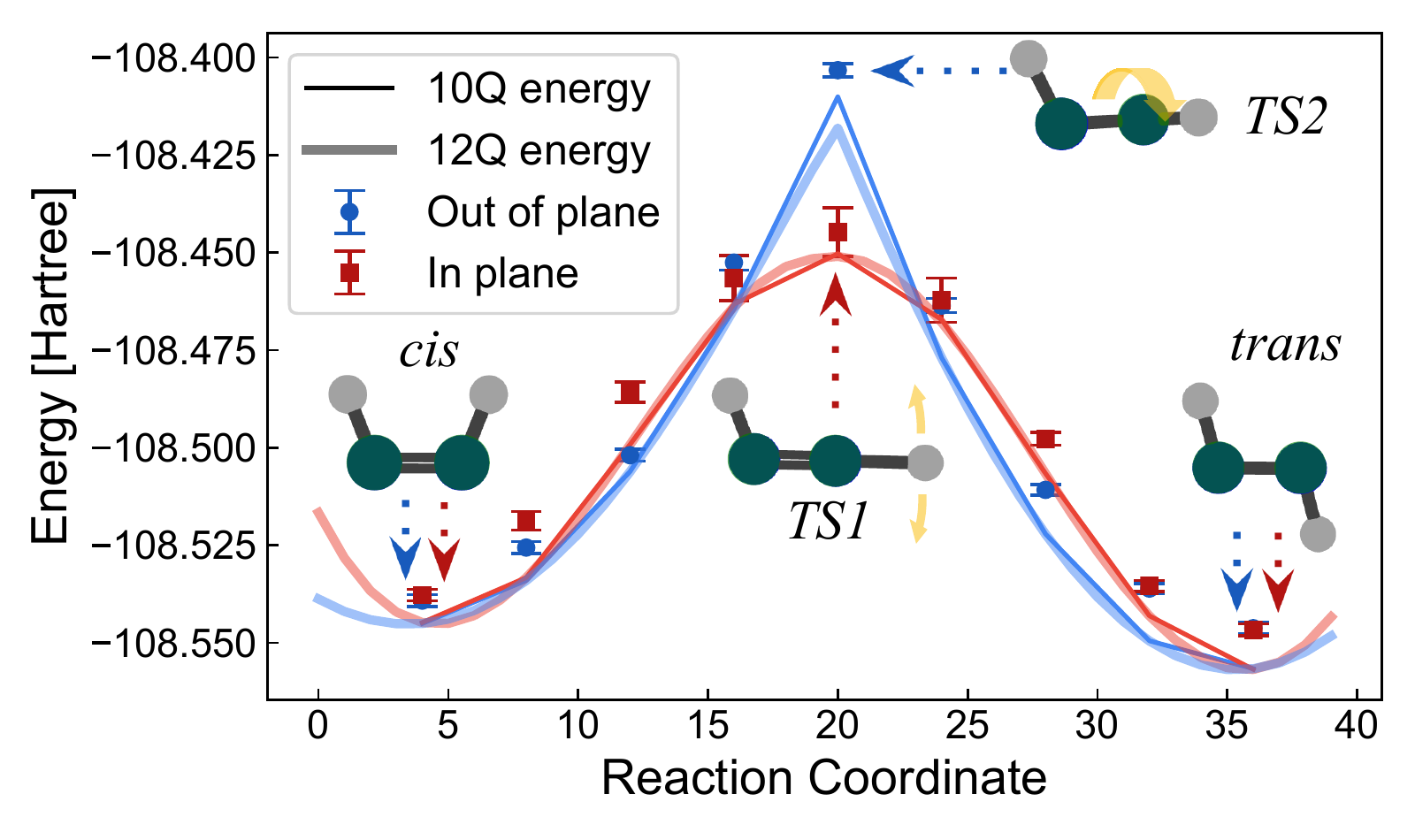}
    \caption{\textbf{VQE performance on distinguishing the mechanism of diazene isomerization.}  Hartree-Fock curves for diazene isomerization between \textit{cis} and \textit{trans} configurations.  $TS1$ and $TS2$ are the transition states for the in-plane and out-of-plane rotation of the hydrogen, respectively. The yellow arrows on $TS1$ and $TS2$ indicate the corresponding reaction coordinate.
    The solid curve is the energy obtained from optimizing a $10$-qubit problem generated by freezing the core orbitals generated from two self-consistent-field cycles.  The transparent lines of the same color are the full $12$ qubit system indicating that freezing the lowest two levels does not change the characteristics of the model chemistry.  Nine points along the reaction paths are simulated on Sycamore using VQE. \nick{We allowed the optimizer 30 iterations for all points except for fifth and sixth point from the left of the in-plane rotation curve which we allowed 60 steps.}  
    \nick{The error bars for all points were computed by estimating the covariance between simultaneously measured sets of $1$-RDM elements and resampling those elements under a multivariate Gaussian model.  Energies from each sample were tabulated and the standard deviation is used as the error bar.  No purification was applied for the computation of the error bar.}
    If purification is applied the error bars become smaller than the markers.  Each basis rotation for diazene contains $50$ $\sqrt{i\textsc{swap}}$ gates and $80$ Rz gates. \label{fig:diazene_simulations}}
\end{figure}

The average energy of any molecular system can be evaluated with knowledge of the one-particle reduced density matrix (1-RDM), $\avg{a^\dagger_p a_q}$, and the two-particle reduced density matrix (2-RDM), $\avg{a^\dagger_p a^\dagger_q a_r a_s}$. In general, it is not possible to exactly reconstruct the 2-RDM from knowledge of just the 1-RDM. However, for single-Slater determinants (as in our Hartree-Fock experiment), the $2$-RDM is completely determined by the $1$-RDM \cite{RevModPhys.32.335}:
\begin{align}
\avg{a^\dagger_p a^\dagger_q a_r a_s} = \avg{a^\dagger_p a_s}\avg{a^\dagger_q a_r} - \avg{a^\dagger_q a_s}\avg{a^\dagger_p a_r}\label{d2_from_d1}.
\end{align}
Thus, in our experiment we only need\nick{ed} to sample the $1$-RDM to estimate the energy. As the $2$-RDM has quadratically more elements than the $1$-RDM, this \nick{approach} is a significant simplification. We measure\nick{d} the $1$-RDM using a protocol described in \app{optimal_rdm_measurement}. This protocol enable\nick{d} us to optimally parallelize measurement of all $N^2$ $1$-RDM elements with $N+1$ distinct circuits. For each distinct circuit we \nick{made} 250,000 measurements.

We perform\nick{ed} two types of error mitigation on our measured data: post-selection on particle number (conserved in basis rotations) and pure-state projection. To apply post-selection we \nick{modified} our circuits by first rotating into a basis that diagonalizes $a^\dagger_p a_q + a^\dagger_q a_p$ for $N$ different pairs of $p$ and $q$ so that these elements \nick{could} be sampled at the same time as the total particle-number operator.  Following the strategy in \app{optimal_rdm_measurement}, this \nick{measurement was} accomplished at the cost of two $T$ gates and one $\sqrt{i\textsc{swap}}$ gate per pair of qubits. We then post-select\nick{ed} to discard measurements where the total number of excitations changed from $\eta/2$.

For pure-state purification, we leverage\nick{d} the fact that the $1$-RDM for any single-Slater determinant wavefunction $\ket{\psi_\kappa}$ has eigenvalues restricted to be 0 and 1~\cite{RevModPhys.35.668}.  We perform\nick{ed} projection back to the pure-set of $1$-RDMs using a technique known as McWeeny purification ~\cite{RevModPhys.32.335}. Details on the procedure and sampling bounds for guaranteeing the procedure has a fixed-point $1$-RDM corresponding to a Slater determinant can be found in \app{purification}.  \nick{Although} McWeeny purification only works for Slater determinant wavefunctions, \nick{pure-state $N$-representability conditions are known for more general systems~\cite{PhysRevA.94.032516} and we expect that a computational procedure similar to enforcing ensemble constraints could be employed ~\cite{rubin2018application,klyachko2006quantum}.} 


A variety of circuit optimization techniques based on gradient- and gradient-free methods have been proposed in the context of NISQ algorithms.  Here, we develop\nick{ed} an optimization technique that exploits local gradient and Hessian information in a fashion which is distinctive to the Hartree-Fock model. It is based on a proposal for iterative construction of a wavefunction to satisfy the Brillouin condition for a single-particle model~\cite{kutzelnigg1979generalized}. Our optimization protocol used the property that at a local optima the commutator of the Hamiltonian $H$ with respect to any generator of rotation $G$ is zero (i.e.~$\bra{\psi} [H, G] \ket{\psi} = 0$) and the fact that sequential basis change circuits can be concatenated into a single basis change circuit (i.e.~$U_a U_b = U_{ab}$). Using these relations and taking $G = \sum_{pq} \kappa_{pq} a^\dagger_p a_q$, as in our experiment, the double commutator $\langle \psi \vert [[H, G], G] \vert \psi \rangle$ determined an augmented Hessian (matrix of derivatives) which we could \nick{use} to iteratively update the wavefunction such that the first order condition \nick{was} approximately satisfied. Regularization \nick{was} added by limiting the size of update parameters~\cite{sun2016co}. For details, see \app{iterative_opt}.  

As a benchmark, we studied symmetrically stretched hydrogen chains of length $6$, $8$, $10$, and $12$ atoms, \fig{fig2_h6_curve}.  \nick{The initial parameters were set to the parameters obtained by solving the Hartree-Fock equations on a classical computer.} The data from the quantum computer is plotted along with classical Hartree-Fock results, showing better and better agreement as we add\nick{ed} post-selection, post-selection and purification, and then error mitigated variational relaxation.  The 6- and 8-qubit data achieved chemical accuracy after VQE, and even the 12-qubit data follow\nick{ed} the expected energy closely.  The error data in \figa{fig2_h6_curve}{b} and the other inserts are remarkable as they show a large and consistent decrease, about a factor of 100, when using these protocols.  \figa{fig2_h6_curve}{c} details the significant decrease in error using a modest number of VQE iterations.  

A fidelity witness can be efficiently computed from the experimental data~\cite{PhysRevLett.120.190501}; see \app{fidelity_measure}.  This value is a lower bound \nick{to the true fidelity}, and thus potentially loose when fidelity is small.  However, \figa{fig2_h6_curve}{b} demonstrates that this fidelity generally tracks the measured errors.  \tab{hydrogen_fidelities} shows how fidelity increase\nick{d} as we add\nick{ed} various forms of error mitigation, starting on the left column where the optimal angles \nick{were} computed classically.   Uncertainties in the last digit, indicated in the parenthesis, are calculated by the procedure described in \app{error_bar_computation}. \nick{The first column of \tab{hydrogen_fidelities} is an estimate of the fidelity based on multiplying the fidelity for all the gates and readout assuming 99.5\% fidelity for single qubit gates, 99\% fidelity for two-qubit gates, and 97\% fidelity for readout.  We see that this estimate qualitatively follows the ``raw'' fidelity witness estimates except when the witness value is very small.}  For all \nick{hydrogen} systems studied, we observed drastic fidelity improvements with combined error mitigation.

\begin{table}[]
    \centering
    \begin{tabular}{|>{\centering\arraybackslash}m{1cm}|>{\centering\arraybackslash}m{1.2cm}|>{\centering\arraybackslash}m{1.2cm}|>{\centering\arraybackslash}m{1.2cm}|>{\centering\arraybackslash}m{1.5cm}|>{\centering\arraybackslash}m{1.5cm}|}
    \hline
    system    & \nick{estimate} &  raw &  +ps & +pure & +VQE \\ \hline
    ${\rm H}_6$ & \nick{0.571} & 0.674(2) & 0.906(2) & 0.9969(1) & 0.99910(9)\\
    ${\rm H}_8$ & \nick{0.412} & 0.464(2) & 0.827(2)& 0.9879(3) & 0.99911(8) \\   
    ${\rm H}_{10}$ & \nick{0.277} & 0.316(2) & 0.784(3) & 0.9704(5) & 0.9834(4)\\
    ${\rm H}_{12}$ & \nick{0.174} & 0.010(2) & 0.654(3) & 0.9424(9) & 0.9913(3)\\
    \hline
    \end{tabular}
    \caption{\textit{\textbf{Average fidelity lower bounds for hydrogen chain calculations}.} We report values of the fidelity witness from \cite{PhysRevLett.120.190501}, averaged across H-H separations of $\{0.5, 0.9, 1.3, 1.7, 2.1, 2.5\} $ \AA, starting from circuits with the theoretically optimal variational parameters ($\kappa$). \nick{``estimate'' corresponds to an estimate of the fidelity derived by multiplying gate errors assuming 0.5 percent single-qubit gate error, 1 percent two-qubit gate error and 3 percent readout error.}  ``Raw'' corresponds to fidelities from constructing the $1$-RDM without any error mitigation.  ``+ps'' corresponds to fidelities from constructing the $1$-RDM with post-selection on particle number. ``+pure'' corresponds to fidelities from  constructing the $1$-RDM with post-selection and applying purification as post-processing.  Finally, ``+VQE'' corresponds to fidelities from  using all previously mentioned error mitigation techniques in conjunction with variational relaxation. Note that for small values (such as the ``raw'' value for $\rm{H}_{12}$) we expect the fidelity lower-bound is more likely to be loose.}
    \label{tab:hydrogen_fidelities}
\end{table}
\textbf{Diazene isomerization.} We simulate\nick{d} two isomerization pathways for diazene, marking the first time that a chemical reaction mechanism has been \nick{modelled} using a quantum computer.  It is known that Hartree-Fock theory reverses the order of the transition states; however, here we focus\nick{ed} on the accuracy of the computation with respect to the simulated model.  Correctly identifying this pathway requires resolving the energy gap of 40 milliHartree between the two \nick{transition} states. The pathways correspond to the motion of the hydrogen in the process of converting \textit{cis}-diazene to \textit{trans}-diazene.  One mechanism is in-plane rotation of a hydrogen \nick{and} the other is an out-of-plane rotation corresponding to rotation of the HNNH dihedral angle.  \fig{diazene_simulations} contains VQE optimized data simulating nine points along the reaction
coordinates for in-plane and out-of-plane rotation of hydrogen. \nick{For all points along the reaction coordinate the initial parameter setting was the solution to the Hartree-Fock equations.} VQE produce\nick{d} $1$-RDMs with average fidelity greater than $0.98$ after error-mitigation. Once again, we see that our full error mitigation procedure significantly improves the accuracy of our calculation.

Our VQE calculations on diazene predict\nick{ed} the correct ordering of the transition states to within the chemical model with an energy gap of $41 \pm 6$ milliHartree \nick{and} the true gap is $40.2$ milliHartree.  We provide a more detailed analysis of the error mitigation performance on the diazene circuits in \app{parasitic_cphase_analysis} considering that the $\sqrt{i\textsc{swap}}$ gates we use\nick{d} \nick{had} a residual $\textsc{cphase}(\pi/24)$ and \nick{Rz gates had stochastic control angles}. This simulation reinforce\nick{d} VQE’s ability to mitigate systematic errors at the scale of 
$50$ $\sqrt{i\textsc{swap}}$ gates and over $80$ Rz gates.


In this work we \nick{took} a step towards answering the question of whether NISQ computers can offer quantum advantage for chemical simulation by studying VQE performance on basis rotation circuits that are widely used in quantum algorithms for fermionic simulation. The \nick{considered} ansatz afford\nick{ed} ways to minimize the resource requirements for VQE and study device performance for circuits that are similar to those needed for full Hamiltonian simulation. These basis rotation circuits also \nick{made} an attractive benchmark due to their prevalence, optimal known compilation, the ability to extract fidelity and fidelity witness values and the fact that they parameterize a continuous family of analytically solvable circuits demonstrating a high degree of entanglement. \nick{The circuits also serve as a natural progression towards more correlated ansatze such as a generalized swap network~\cite{Kivlichan2017} or a non-particle conserving circuit ansatz followed by particle number projection.}

We demonstrated the performance of two error mitigation techniques on basis rotation circuit fidelity.  The first is post-selection on total occupation number when measuring all elements of the $1$-RDM.  This \nick{step was} accomplished by permuting the basis rotation circuit such that all measurements involve\nick{d} estimating nearest-neighbor observables and measuring each pair of observables such that the total occupation number is preserved.  The second is the application of McWeeny purification as a post-processing step.  The energy improvements \nick{from} projecting back to the pure-state $N$-representable manifold \nick{was evidence} that generalized pure-state $N$-representability conditions \nick{would} be instrumental in making NISQ chemistry computations feasible. This \nick{fact} underscores the importance of developing procedures for applying pure-state $N$-representability conditions in a more general context.  \nick{The post-selection and RDM measurement techniques can be generalized to measuring all $1$-RDM and $2$-RDM elements when considering a less restrictive circuit ansatz by permuting the labels of the fermionic modes.  For ansatz such as the generalized swap network~\cite{Kivlichan2017} the circuit structure would not change, only the rotation angles. Thus, the measurement schemes presented here are applicable in the more general case.} \nick{Furthermore, it is important to understand the performance of these error mitigation techniques when combined with alternatives such as noise extrapolation~\cite{PhysRevLett.119.180509}}.   

Finally, we were able to show further evidence that variational relaxation effectively mitigate\nick{s} coherent errors arising in implementation of physical gates.  \nick{The performance of our problem specific optimization strategy motivates the study of iterative wavefunction constructions~\cite{grimsley2019adaptive} in a more general setting.} The combination of these error mitigation techniques with VQE unambiguously resolve\nick{d} a chemical mechanism to within the model chemistry using a quantum computation. 
It is still an open question whether NISQ devices will be able to simulate challenging quantum chemistry systems and it is likely that major innovations would be required.
However, we find the accuracy of these experiments and the effectiveness of these error-mitigation procedures to be an encouraging signal of progress in that direction.


\section*{Acknowledgements}
D.B. is a CIFAR Associate Fellow in the Quantum Information Science Program.
\textbf{Funding:} This work was supported by Google. 
\textbf{Competing Interests:}
\nick{The authors declare no competing interests.}
\textbf{Author Contributions:}
N.C.R. designed the experiment. C.N. assisted with data collection. Z.J., V.S., and N.W. assisted with analytical calculations and gate synthesis. N.C.R. and R.B. wrote the paper. Experiments were performed using a quantum processor that was recently developed and fabricated by a large effort involving the entire Google Quantum team.
\textbf{Data and materials availability:}
The code used for this experiment and a tutorial for running it can be found in the open source library Recirq, located at \url{https://github.com/quantumlib/ReCirq/tree/master/recirq}.  All data needed to evaluate the conclusions in the paper are present in the paper or the Supplementary Materials.  Data presented in the figures can be found in the Dryad repository located at~\cite{dryadrepo}

\newpage
\onecolumngrid

\vspace{1em}
\begin{flushleft}
{\Large Google AI Quantum and Collaborators}

\bigskip

\renewcommand{\author}[2]{#1$^\textrm{\scriptsize #2}$}
\renewcommand{\affiliation}[2]{$^\textrm{\scriptsize #1}$ #2 \\}
\newcommand{\xGoogle}{\affiliation{1}{Google Research}}
\newcommand{\xUMass}{\affiliation{2}{Department of Electrical and Computer Engineering, University of Massachusetts, Amherst, MA}}
\newcommand{\xUCSB}{\affiliation{3}{Department of Physics, University of California, Santa Barbara, CA}}
\newcommand{\xUCB}{\affiliation{4}{Department of Chemistry, University of California, Berkeley, CA}}
\newcommand{\xUCR}{\affiliation{5}{Department of Electrical and Computer Engineering, University of California, Riverside, CA}}
\newcommand{\xUM}{\affiliation{6}{Department of Electrical Engineering and Computer Science, University of Michigan, Ann Arbor, MI}}
\newcommand{\xDaimler}{\affiliation{7}{Mercedes-Benz Research and Development, North America, Sunnyvale, CA}}
\newcommand{\xUW}{\affiliation{8}{Department of Physics, University of Washington, Seattle, WA}}
\newcommand{\xPNNL}{\affiliation{9}{Pacific Northwest National Laboratory, Richland, WA}}

\newcommand{\Google}{1}
\newcommand{\UMass}{2}
\newcommand{\UCSB}{3}
\newcommand{\UCB}{4}
\newcommand{\UCR}{5}
\newcommand{\UM}{6}
\newcommand{\Daimler}{7}
\newcommand{\UW}{8}
\newcommand{\PNNL}{9}

\author{Frank Arute}{\Google},
\author{Kunal Arya}{\Google},
\author{Ryan Babbush}{\Google},
\author{Dave Bacon}{\Google},
\author{Joseph C.~Bardin}{\Google,\! \UMass},
\author{Rami Barends}{\Google},
\author{Sergio Boixo}{\Google},
\author{Michael Broughton}{\Google},
\author{Bob B.~Buckley}{\Google},
\author{David A.~Buell}{\Google},
\author{Brian Burkett}{\Google},
\author{Nicholas Bushnell}{\Google},
\author{Yu Chen}{\Google},
\author{Zijun Chen}{\Google},
\author{Benjamin Chiaro}{\Google,\! \UCSB},
\author{Roberto Collins}{\Google},
\author{William Courtney}{\Google},
\author{Sean Demura}{\Google},
\author{Andrew Dunsworth}{\Google},
\author{Daniel Eppens}{\Google},
\author{Edward Farhi}{\Google},
\author{Austin Fowler}{\Google},
\author{Brooks Foxen}{\Google},
\author{Craig Gidney}{\Google},
\author{Marissa Giustina}{\Google},
\author{Rob Graff}{\Google},
\author{Steve Habegger}{\Google},
\author{Matthew P.~Harrigan}{\Google},
\author{Alan Ho}{\Google},
\author{Sabrina Hong}{\Google},
\author{Trent Huang}{\Google},
\author{William J. Huggins}{\Google,\! \UCB},
\author{Lev Ioffe}{\Google},
\author{Sergei V.~Isakov}{\Google},
\author{Evan Jeffrey}{\Google},
\author{Zhang Jiang}{\Google},
\author{Cody Jones}{\Google},
\author{Dvir Kafri}{\Google},
\author{Kostyantyn Kechedzhi}{\Google},
\author{Julian Kelly}{\Google},
\author{Seon Kim}{\Google},
\author{Paul V.~Klimov}{\Google},
\author{Alexander Korotkov}{\Google,\! \UCR},
\author{Fedor Kostritsa}{\Google},
\author{David Landhuis}{\Google},
\author{Pavel Laptev}{\Google},
\author{Mike Lindmark}{\Google},
\author{Erik Lucero}{\Google},
\author{Orion Martin}{\Google},
\author{John M.~Martinis}{\Google,\! \UCSB},
\author{Jarrod R.~McClean}{\Google},
\author{Matt McEwen}{\Google,\! \UCSB},
\author{Anthony Megrant}{\Google},
\author{Xiao Mi}{\Google},
\author{Masoud Mohseni}{\Google},
\author{Wojciech Mruczkiewicz}{\Google},
\author{Josh Mutus}{\Google},
\author{Ofer Naaman}{\Google},
\author{Matthew Neeley}{\Google},
\author{Charles Neill}{\Google},
\author{Hartmut Neven}{\Google},
\author{Murphy Yuezhen Niu}{\Google},
\author{Thomas E.~O'Brien}{\Google},
\author{Eric Ostby}{\Google},
\author{Andre Petukhov}{\Google},
\author{Harald Putterman}{\Google},
\author{Chris Quintana}{\Google},
\author{Pedram Roushan}{\Google},
\author{Nicholas C.~Rubin}{\Google},
\author{Daniel Sank}{\Google},
\author{Kevin J.~Satzinger}{\Google},
\author{Vadim Smelyanskiy}{\Google},
\author{Doug Strain}{\Google},
\author{Kevin J.~Sung}{\Google,\! \UM},
\author{Marco Szalay}{\Google},
\author{Tyler Y. Takeshita}{\Daimler},
\author{Amit Vainsencher}{\Google},
\author{Theodore White}{\Google},
\author{Nathan Wiebe}{\Google,\! \UW,\! \PNNL},
\author{Z.~Jamie Yao}{\Google},
\author{Ping Yeh}{\Google},
\author{Adam Zalcman}{\Google}

\bigskip

\xGoogle
\xUMass
\xUCSB
\xUCB
\xUCR
\xUM
\xDaimler
\xUW
\xPNNL

\end{flushleft}

\twocolumngrid
\bibliography{biblo,ryan}
\onecolumngrid
\newpage

\appendix

\section{Hartree-Fock Theory via Canonical Transformations}\label{app:hf_via_ct}
In this section we derive Hartree-Fock theory from the perspective of canonical transformations.  This derivation follows an original work by David Thouless~\cite{Thouless1960} and is reproduced here due to its foundational importance to the formulation of this experiment. In Hartree-Fock theory one attempts to solve the time-independent Schr\"odinger equation using a state ansatz that is an \nick{antisymmetrized} product of one-particle functions.   
Starting from an arbitrary orthogonal basis $\{\phi_{i}\}$ the goal is to variationally optimize the wavefunction
\begin{align}
\vert \psi(r_{1}, ..., r_{n}) \rangle = (n!)^{-1/2} A_{n}\left(\chi_{1}(r_{1})...\chi_{n}(r_{n})\right)
\end{align}
where $A_{n}$ is the \nick{antisymmetrizer} and $\chi_{i}(r) = \sum_{j}c_{i}^{j}\phi_{j}(r)$ in terms of the coefficients for $\chi$.   This \nick{antisymmetrized} product of one-particle functions is commonly expressed in a more compact form as a determinant of a matrix whose elements are the functions $\chi_{i}(r_{j})$ with $i$ indexing the column and $j$ indexing the row of this matrix.  This representation of the \nick{antisymmetrized} product through a determinant is why this wavefunction ansatz is commonly referred to as a Slater determinant. 

The variational principle for the Schr\"odinger equation can be stated as 
\begin{align}\label{eq:first_order_variation}
\langle \delta \psi \vert H \vert \psi \rangle = 0
\end{align}
which is a statement that the energy is stationary with respect to first order changes in the wavefunction.    In second quantization a single \nick{antisymmetrized} product of orbitals corresponds to a product of ladder operators acting on the vacuum to ``create'' a representation of the \nick{antisymmetrized} wavefunction
\begin{align}
\langle r \vert \psi \rangle =& \langle r \vert \prod_{i=1}^{n}a_{i}^{\dagger} \vert 0 \rangle
                   = \frac{1}{\sqrt{n!}}\mathrm{Det}\left[
\begin{pmatrix}
\chi_{1}(r_{1}) & ... & \chi_{1}(r_{n}) \\
\vdots & \ddots & \vdots \\
\chi_{n}(r_{1}) & ... & \chi_{n}(r_{n})
\end{pmatrix}\right].
\end{align}
Assuming we are working in a fixed particle manifold and given the aforementioned complete set of one-particle functions is used as a basis we can index the functions used in the product wavefunction by $i$ and those not used are labeled by $a$ then any change in the wavefunction is generated by
\begin{align}
\langle \delta \psi\vert = \langle \psi \vert a_{i}^{\dagger}a_{a} \zeta
\end{align}
where $\zeta$ is the first order change to an orbital $\chi_{i}$.
This fact is because any unitary generator that has only indices $\{a\}$ or $\{i\}$ merely changes the phase on the state and thus is not observable~\cite{szabo2012modern}.  Evaluating \eq{first_order_variation} one arrives at an expression for the stationarity of the state
\begin{align}\label{eq:hamiltonian_stopping_condition}
\langle \psi \vert a_{i}^{\dagger}a_{a} H \vert \psi \rangle = 0.
\end{align}
All the quantities in \eq{hamiltonian_stopping_condition} can be evaluated using Wick's theorem given the initial state $\psi$ is a product state and $a_{r}\vert 0 \rangle = 0$.  This variational condition naturally leads to the \textit{self-consistent-field} Hamiltonian commonly derived through a Lagrangian technique for the Hartree-Fock equations.  In order to design a VQE style approach to solving the Hartree-Fock equations we take a different approach that  leverages the fact that we can determine any basis rotation through a linear-depth quantum circuit.  
Thouless demonstrated~\cite{Thouless1960} that any non-orthogonal product wavefunction can be obtained from a product wavefunction by a unitary generated by one-body fermionic operators of the form $a_{p}^{\dagger}a_{q}$. The underlying reason for why this \nick{fact} is true is that the one-body fermionic generators form a closed Lie-algebra.  
Given,
\begin{align}
\left[a_{p}^{\dagger}a_{q}, a_{r}^{\dagger}a_{s}\right] = \delta_{q, r}a_{p}^{\dagger}a_{s} - \delta_{p, s}a_{r}^{\dagger}a_{s}
\end{align}
the adjoint representation of any element of the algebra $\kappa$ where
\begin{align}
\kappa = \sum_{p,q}\kappa_{p, q}a_{p}^{\dagger}a_{q},
\end{align}
and its commutator with any other element can be efficiently represented as matrix that is $m \times m$ where $m$ is the number of fermionic modes.  
\begin{align}
\left[ \kappa, a_{p}^{\dagger} \right] = a_{q}^{\dagger}\kappa_{p, q} \;\;,\;\;\left[ \kappa, a_{p} \right] = a_{q}\kappa_{p, q}^{*}
\end{align}

Using the BCH expansion, we can express the similarity transformed ladder operators as
\begin{align}\label{eq:similarity_one_body}
e^{K}a_{p}^{\dagger}e^{-K} = \sum_{q}a_{q}^{\dagger}u_{q, p} \;\;,\;\;e^{K}a_{p}e^{-K} = \sum_{q}a_{q}u_{q, p}^{*}
\end{align}
where $u$ is the matrix given by the exponentiation of the coefficient matrix for the generator operator $\kappa$
\begin{align}\label{eq:slater_determinant_orbitals}
u = e^{\kappa}
\end{align}
which is the proof for \eq{basis_change}. Any rotation of the underlying basis can now be represented as a similarity transformation of each fermionic mode
\begin{align}\label{eq:basis_rotation_ansatz}
\vert \phi(\kappa) \rangle = e^{K}a_{1}^{\dagger}e^{-K}...e^{K}a_{n}^{\dagger}e^{-K}\vert 0 \rangle
= e^{K} \vert \psi \rangle.
\end{align}
Thus any non-orthogonal state can be generated by implement\nick{ing} $e^{K}$ as a circuit acting on an initial product state.  

Given the Hartree-Fock wavefunction ansatz the energy is given by 
\begin{align}
E(\kappa) = \langle \phi(\kappa) \vert H\vert \phi(\kappa)\rangle
\label{eq:energy_hf}
=\langle \psi\vert e^{K}  H e^{-K}\vert \psi\rangle.
\end{align}
With the energy expressed in the form of \eq{energy_hf} it is not immediately clear that it can be evaluated without knowledge of the $2$-RDM.  To see this \nick{fact}, \nick{we used the} BCH expansion and notice that all nested commutator terms involve $a_{p}^{\dagger}a_{q}$-like terms and the original Hamiltonian.  The commutator of a two-mode number conserving fermionic operator with a four-mode number conserving fermionic operator produces a linear combination of four four-mode number conserving fermionic operators.  Therefore, all terms in the expansion can be evaluated with knowledge of only the $2$-RDM.  If we start with a product state defined from an orthogonal set of states the $2$-RDM can be constructed directly from the $1$-RDM~\cite{RevModPhys.32.335}
\begin{align}
{}^{1}D_{i}^{j} =& \langle\phi \vert a_{j}^{\dagger}a_{i}\vert \phi \rangle \nonumber \\
{}^{2}D_{ij}^{pq} =& \langle \phi \vert a_{p}^{\dagger}a_{q}^{\dagger}a_{j}a_{i} \vert \phi \rangle
= {}^{1}D_{i}^{p}~{}^{1}D_{j}^{q} - {}^{1}D_{i}^{q}~{}^{1}D_{j}^{p} \label{eq:d2_from_d1}.
\end{align}
This \nick{expression} also demonstrates that we only need to measure the $1$-RDM to evaluate the energy. \nick{It is important to note that the reconstruction of the 2-RDM from the 1-RDM described in Eq.~\eqref{d2_from_d1} is only exact for Slater determinant wavefunctions.} The energy is evaluated as a function of the $1$- and $2$-RDM by
\begin{align}
E(\kappa) = \sum_{ij}h_{ij}\langle \phi(\kappa) \vert a_{i}^{\dagger}a_{j} \vert \phi(\kappa) \rangle 
+ \sum_{ijkl}V_{ijkl}\langle \phi(\kappa) \vert a_{i}^{\dagger}a_{j}^{\dagger}a_{k}a_{l}\vert \phi(\kappa) \rangle
= \sum_{ij}h_{ij}{}^{1}D_{j}^{i} + \sum_{ijkl}V_{ijkl}{}^{2}D_{lk}^{ij} \label{eq:energy_from_d1_d2}
\end{align}
where $h_{ij}$ and $V_{ijkl}$ 
\begin{align}
h_{i,j} =& \int dr \chi_{i}^{*}(r)\left(-\nabla^{2}(r) + \sum_{A}\frac{Z_{A}}{|r - R_{A}|}\right) \chi_{j}(r) \\
V_{l,k}^{i,j} =& \frac{1}{2}\int \int dr_{1}dr_{2} \chi_{i}^{*}(r_{1})\chi_{j}^{*}(r_{2})\left(|r_{1} - r_{2}|^{-1}\right)\chi_{k}(r_{2})\chi_{l}(r_{1})
\end{align}
are the molecular integrals in the original basis.  These orbitals are determined by diagonalizing the matrix of one-body integrals $h_{ij} = [\mathbf{h}]_{ij}$ described in the \nick{STO-3G} atomic basis. 
In summary, to measure the energy of our system given basis rotation circuit ansatz we need the following steps:
\begin{enumerate}
\item[1] Measure the entire $1$-RDM.
\item[2] Compute the $2$-RDM by evaluating. \eq{d2_from_d1}
\item[3] Compute the energy by evaluating. \eq{energy_from_d1_d2}
\end{enumerate}

\subsection{Classical simulation of non-interacting fermion circuits}\label{app:classical_sim_of_free_fermions}
Given a particular set of parameters $\{\kappa_{p, q}\}$ the $1$-RDM resulting from a wavefunction $\psi = U(\kappa)\phi$, where $\phi$ is an initial product state, is
\begin{align}
{}^{1}\tilde{D}_{q}^{p} =& \langle \phi \vert e^{-K} a_{p}^{\dagger} e^{K}e^{-K}a_{p} e^{K} \vert \phi \rangle 
= \langle \phi \vert \sum_{j}u_{p, i}a_{i}^{\dagger} \sum_{q, j}u_{q, j}^{*}a_{j}  \vert \phi \rangle 
= \sum_{ij}u_{p, i}u_{q, j}^{*}\langle \phi \vert a_{i}^{\dagger} a_{j}  \vert \phi \rangle.
\end{align}
With this $1$-RDM one can evaluate the energy and gradients with respect to $\kappa_{p,q}$.  This expression requires two matrix multiplications to evaluate along with the $1$-RDM of the starting state.  

\section{Implementing the Basis Change Circuit and Circuit Concatenation}\label{app:optimal_compilation}

In order to implement the basis rotation circuits we leverage a number of recent works that provide asymptotically optimal circuit compilations.  We review a circuit construction that is analogous to a QR decomposition as motivation before highlighting the salient features of the optimal circuit compilation.  The basis rotation circuit is first expressed in fermionic modes which we then provide a compilation to the gate set used in this work.   Our goal is to implement a unitary corresponding to
\begin{align}
U(e^{\kappa}) = e^{K} \qquad \qquad
K = \sum_{i, j}\kappa_{i, j}a_{i}^{\dagger}a_{j} \qquad \qquad
\mathbf{\kappa}^{\dagger} = -\mathbf{\kappa}.
\end{align}
Not all terms in $K$ commute and thus naively one would expect an approximate method such as Trotterization to be required.  In Reference~\cite{Kivlichan2017} the connection of the QR decomposition of $e^{\kappa}$ via Givens rotation to the sequence of untiaries $R_{pq}(u)$
\begin{align}\label{eq:nn_reck_unitary}
R(u)_{pq} = e^{\mathrm{log}\left[u\right]_{pq}(a_{p}^{\dagger}a_{q}-a_{q}^{\dagger}a_{p})}
\end{align}
was established allowing for the exact evolution of the one-body component of the Hamiltonian without Trotter error and a circuit to implement any basis rotation--i.e. any Slater determinant.  
A \nick{distinctive} feature of one-body rotations is that the map $U(e^{\kappa})$ is a homomorphism under matrix multiplication
\begin{align}\label{eq:one_body_homomorphism}
U(e^{\kappa}) \cdot U(e^{\kappa'}) = U(e^{\kappa} \cdot e^{\kappa'})
\end{align}
We use this homomorphism through the observation that
\begin{align}
R_{pq}(\theta)U(u) = U(r_{pq}(\theta)u)
\end{align}
where
\begin{align}
r(\theta)_{p, q} = 
\begin{pmatrix}
1 & ... & 0 & ... & 0 & ... & 0 \\
\vdots & \ddots & \vdots &  & \vdots &  & \vdots \\
0 & ... & \cos(\theta) & ... & -\sin(\theta) & ... & 0\\
\vdots &   & \vdots &  & \vdots &  & \vdots \\
0 & ... & \sin(\theta) & ... & \cos(\theta) & ... & 0\\
\vdots &  & \vdots &  & \vdots & \ddots & \vdots \\
0 & ... & 0 & ... & 0 & ... & 1 \\
\end{pmatrix}
\end{align}
which given an appropriate selection of a sequence of $r_{p,q}(\theta)$ brings $u$ into diagonal form
\begin{align}
\prod_{k}R_{k}(\theta_{k})U(u) =& \sum_{p}e^{-i\phi_{p}a_{p}^{\dagger}a_{p}} = \sum_{p}e^{-i\phi_{p}\vert p \rangle \langle p\vert}
\end{align}
The sequence of $R_{k}(\theta_{k})$ can be determined by a QR decomposition of the matrix $u$.   This \nick{fact} was first recognized by Reck~\cite{PhysRevLett.73.58} and used in a variety of quantum optics experiments to implement universal unitary operations--limited to unitaries associated with one-body fermionic Hamiltonians. Jiang \textit{et. al} and Kivlichan \textit{et. al}~\cite{PhysRevApplied.9.044036, Kivlichan2017} point out that in a fixed particle manifold the circuit depth can be further minimized.  This \nick{fact} is clearly shown by considering the state in the basis that is being prepared through the Givens rotation network and back transforming to the original basis
\begin{align}
\vert \psi(\kappa)\rangle =& \prod_{i=1}^{\eta}\tilde{a}_{i}^{\dagger}\vert \mathrm{vac} \rangle = \prod_{i=1}^{\eta}e^{-K}\tilde{a}_{i}^{\dagger}e^{K}\vert \mathrm{vac} \rangle  = \prod_{i=1}^{\eta}\sum_{p}\left[e^{\kappa}\right]_{i,p}a_{p}^{\dagger}\vert \mathrm{vac} \rangle 
\end{align}
we only need the first $\eta$-columns of the matrix $\left[e^{\kappa}\right]$.  Therefore, we can focus on Givens network elimination on these columns.  Jiang \textit{et al.} provide a further circuit minimization by noting that any rotation amongst the occupied orbitals merely shifts the observable by a global phase.  Given a unitary $V$
\begin{align}
\prod_{i=1}^{\eta}\sum_{j=1}^{\eta}V_{i,j}\tilde{a}_{i}^{\dagger}\vert \mathrm{vac} \rangle = \mathrm{det}\left[V\right] \prod_{i=1}^{\eta}\tilde{a}_{i}^{\dagger}\vert \mathrm{vac} \rangle
\end{align}
where the $\mathrm{det}\left[V\right]$ is a phase and thus not observable.  The $V$ can be chosen such that the lower left triangle or $e^{\kappa}$ are zeroed out by Givens rotations.  In chemistry parlance, \nick{this transformation is} called an occupied-occupied orbital rotations and is known to be a redundant rotation.  For restricted Hartree-Fock the number of non-redundant parameters in $\kappa$ is equal to the number of occupied spatial orbitals times the number of virtual orbitals.  We also note that this decomposition is exact and likely asymptotically optimal.  \nick{While the authors of~\cite{Kivlichan2017} argue that in terms of gate count their Givens rotation circuits are likely optimal, we note that approximate unitary constructions such as those in ~\cite{daskin2011decomposition} may provide a route to approximating the compilation of similar circuits with even fewer gates.}

\begin{figure*}[h]
    \centering
    \includegraphics[width=\linewidth]{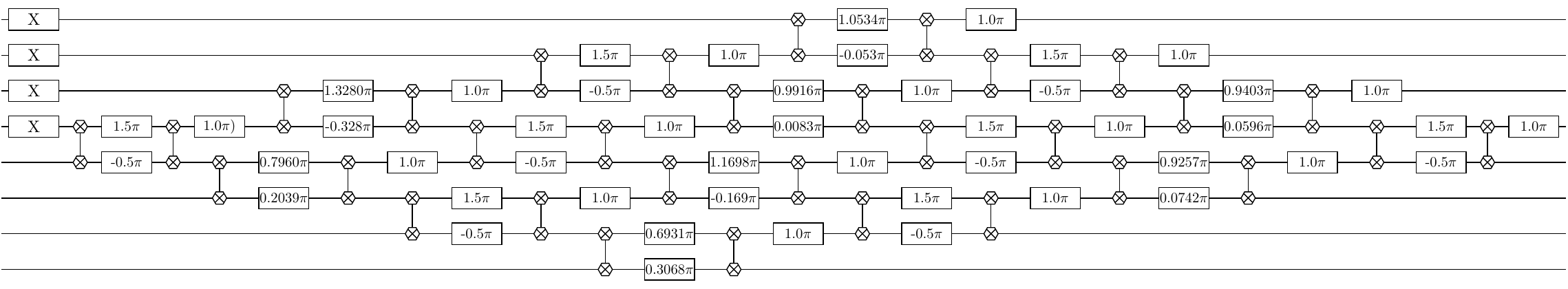}
    \caption{Givens rotation circuit for H$_{8}$ simulating a random basis transformation in the half filling sector.  Each Givens rotation is compiled into $\sqrt{i\textsc{swap}}$ \nick{(hexagon two-qubit gates)} and $\mathrm{Rz}$ gates \nick{(square gates with an angle depicted)}.}
    \label{fig:h8_circuit}
\end{figure*}
\begin{figure*}[h]
    \centering
    \includegraphics[width=\linewidth]{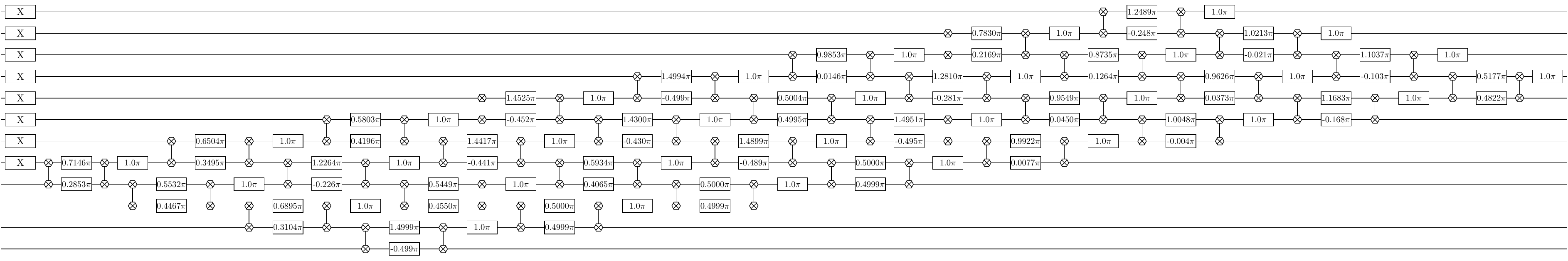}
    \caption{Givens rotation circuit for diazene prior to freezing the two lowest energy orbitals.  Away from half filling the basis rotations have a parallelogram structure.}
    \label{fig:diazene_circuit}
\end{figure*}

An example of an eight qubit half-filling circuit is given in \figs{h8_circuit}. When we are away from half filling the nice symmetry of the circuit is lost.  For example, \figs{diazene_circuit} is Diazene which has 8-electrons in 12 orbitals.

\section{Optimal Measurement of the 1-RDM}\label{app:optimal_rdm_measurement}
In this section we present a methodology that allows us to measure the $1$-RDM in $N+1$ measurement settings and  no additional quantum resources.  We will also discuss a method that allows us to perform post selection on all the Monte Carlo averaged terms at the cost of an additional row of $\sqrt{i\textsc{swap}}$ gates at the end of the circuit. The $1$-RDM is an $N\times N$ hermitian positive semidefinite matrix with elements equal to the expectation values $\langle a_{i}^{\dagger}a_{j} \rangle$ where $\{i, j\}$ index the row and column of the matrix. The matrix of expectation values is depicted in \figs{diag_opdm}.  As a motivator for our measurement protocol we start by describing circuits required to measure the diagonal elements of the $1$-RDM of a six qubit system at half filling--i.e. $\langle a_{i}^{\dagger}a_{i}\rangle$.
\subsection{Diagonal terms}\label{sub_sec:diagonal_terms}
\begin{figure}[H]
    \centering
    \includegraphics[width=10cm]{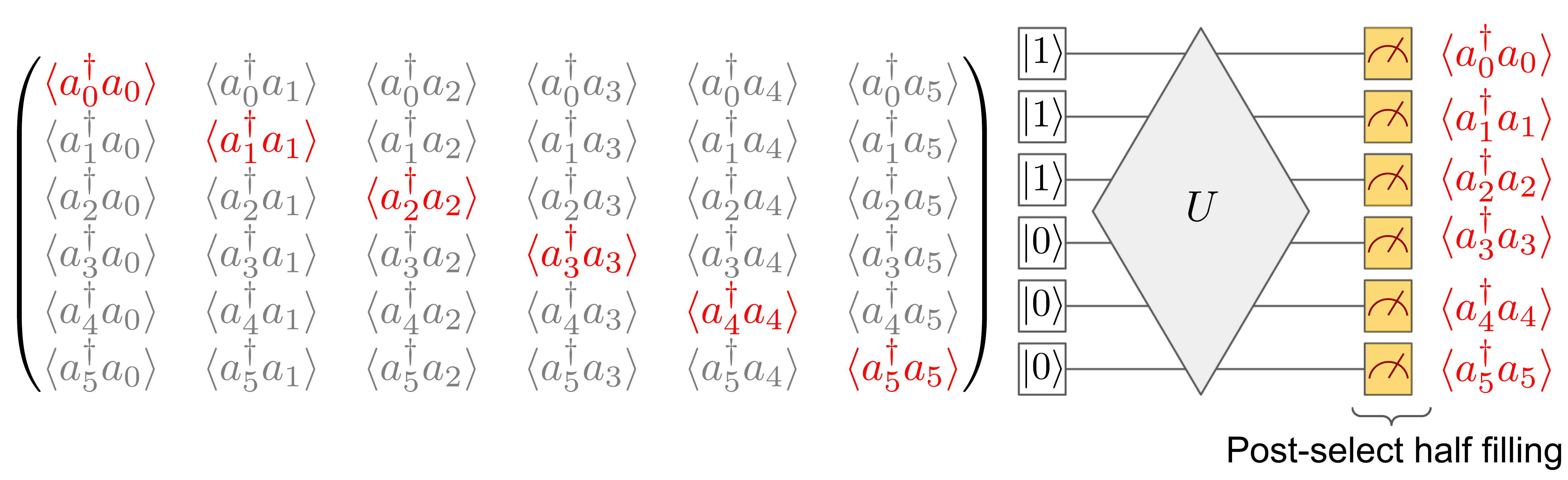}
    \caption{Measurement circuit associated with estimating all diagonal elements of the $1$-RDM simultaneously.  The elements that are acquired with this circuit are highlighted in red.}
    \label{fig:diag_opdm}
\end{figure}

Given a circuit $U$ implementing the basis rotation $e^{\kappa}$ the diagonal elements of the $1$-RDM are obtained by measuring the $Z$ expectation value on each qubit. The correspondence between $a_{i}^{\dagger}a_{i}$, measurement result $M_{i}$ from qubit $i$, qubit operators is derived using the Jordan-Wigner transform
\begin{align}
\langle a_{i}^{\dagger}a_{i}\rangle = \frac{I - \langle Z_{i}\rangle}{2} = \langle M_{i} \rangle
\end{align}
where $Z_{i}$ is the $Z$-qubit operator on qubit labeled $i$.  The expectation value $\langle a_{i}^{\dagger}a_{i}\rangle$ is equivalent to the probability of measuring a $1$ bit on qubit $i$--i.e $\langle M_{i}\rangle$.  Because we are measuring in the computational basis we can post-select on the three excitations in the measurement result.  This process is depicted in \figs{diag_opdm}.
\subsection{One-off-diagonal terms}\label{sub_sec:one_off_diagonal_terms}
\begin{figure}[H]
    \centering
    \includegraphics[width=12.5cm]{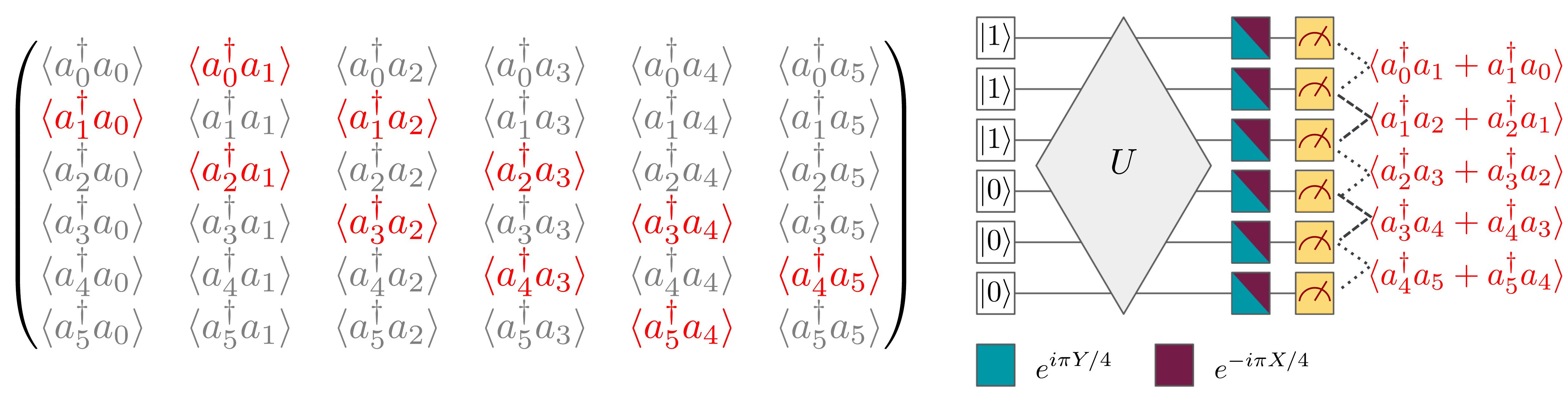}
    \caption{The two circuits allowing for the measurement of all one-off-diagonal elements of the $1$-RDM simultaneously. The teal circuit involves performing an Ry rotation (to measure in the $X$ basis) at the end of the circuit \nick{and} the purple circuit contains an Rx rotation (to measure in the $Y$ basis).  The 1-RDM elements that are acquired with these circuits are highlighted in red.  We label which pairs contribute to which expectation values with grey dashed lines.  The thinner dashes are for the even $1$-RDM pairs \nick{and} the thicker dashes are for the odd $1$-RDM pairs.  Because Ry and Rx operations do not preserve particle number we cannot post-select on total particle number with these measurement circuits.}
    \label{fig:k1_opdm}
\end{figure}

The hermiticity of the $1$-RDM demands that $\langle a_{i}^{\dagger}a_{i+1} \rangle = \langle a_{i+1}^{\dagger}a_{i}\rangle^{*}$. The $1$-RDM has no imaginary component because we use an initial basis built from real valued orbitals and the basis rotation circuit implements an element of $SO(N)$--i.e. the basis rotation circuit involves a unitary matrix with real values. Therefore, we only measure the real part of all one-off-diagonal terms $a_{i}^{\dagger}a_{i+1} + a_{i+1}^{\dagger}a_{i}$ which corresponds to $2 \Re{\langle a_{i}^{\dagger}a_{i+1}\rangle}$.  Using the Jordan-Wigner transform to map fermionic ladder operators to qubits
\begin{align}
\langle a_{i}^{\dagger}a_{i+1} + a_{i+1}^{\dagger}a_{i}\rangle = \frac{1}{2}\left(\langle X_{i}X_{i+1}\rangle + \langle Y_{i}Y_{i+1}\rangle\right) = 2 \Re{\langle a_{i}^{\dagger}a_{i+1}\rangle}
\end{align}
we see that we must measure $XX$ on all pairs and $YY$ on all pairs.  This \nick{measurement} can be accomplished with two circuits depicted in \figs{k1_opdm}.
\subsection{General off-diagonal terms and virtual swapping}
The label of each fermionic mode is an arbitrary choice, so we are free to reorder the labels such that measuring nearest-neighbor pairs of qubits corresponds to measuring different off-diagonal $1$-RDM elements.  Every relabeling of the qubits requires us to recompile the Givens rotation circuit.  The structure of the circuit stays the same but the rotation angles are different.  In this section we describe how to recompute the Givens rotation angles based on a new label ordering.  Using the label sets $\{1, 3, 0, 5, 2, 4\}$ and $\{3, 5, 1, 4, 0, 2\}$ we are able to use the two \nick{measurement} circuits in \figs{k1_opdm} to measure the remaining off-diagonal $1$-RDM elements.

Formally, we build the new qubit labels by \nick{virtually} swapping fermionic modes at the end of the original circuit implement $e^{\kappa}$.  
We note that performing nearest-neighbor fermionic swaps between adjacent pairs twice (even swaps and odd swaps) we obtain a new ordering of qubits. 
For example, consider six fermionic modes $\{0, 1, 2, 3, 4, 5\}$. Performing a set of fermionic swaps on modes labeled $\{(0, 1), (2, 3), (4, 5)\}$ followed by swaps on $\{(1, 2), (3, 4)\}$ leaves our mode ordering as $\{1, 3, 0, 5, 2, 4\}$.  We can then perform $X$-Pauli and $Y$-Pauli measurements on each qubit to recover expectation values associated with
\begin{align}
\{\Re(a_{1}^{\dagger}a_{3} + a_{3}^{\dagger}a_{1}), \Re(a_{3}^{\dagger}a_{0} + a_{0}^{\dagger}a_{3}),
\Re(a_{0}^{\dagger}a_{5} + a_{5}^{\dagger}a_{0}), \Re(a_{5}^{\dagger}a_{2} + a_{2}^{\dagger}a_{5}),
\Re(a_{2}^{\dagger}a_{4} + a_{4}^{\dagger}a_{2}) \}.
\end{align}
This procedure can be repeated once more to measure all the required two-body fermionic correlators to construct the $1$-RDM.  Though it appears that each new label set incurs additional circuit by requiring fermionic swaps between neighboring modes we can exploit the fact that one-body fermionic swaps generated by $\mathrm{exp}(-i \pi \textsc{fswap} / 2)$ where $\textsc{fswap}$ is
\begin{align}
\textsc{fswap} = a_{p}^{\dagger}a_{q} + a_{q}^{\dagger}a_{p} - a_{p}^{\dagger}a_{p} - a_{q}^{\dagger}a_{q}.
\end{align}
This one-body permutation can be viewed as a basis rotation which can be concatenated with the original circuit at no extra cost due to \eq{one_body_homomorphism}.  The swapping unitary simply shuffles the columns of $e^{\kappa}$ that is used to generate the Givens rotation network.  \nick{The same effect could have been achieved by relabeling the fermionic modes which would have been equivalent to permuting the rows and columns of $e^{\kappa}$.  This relabeling technique can be applied beyond basis rotation circuits.  For example, one can relabel the fermionic modes of a generalized swap network such that different sets of RDM elements can be measured as nearest-neighbor pairs.  The same logic can be applied to $k$-RDM elements.}

In conclusion we need $N/2$ circuits, where each of the $N/2$ circuits gets measured in two or three different ways, for an $N$-qubit system to measure the $1$-RDM.  \nick{This is a quadratic improvement over the naive measurement scheme which would require ${\cal O}(N^2)$ different measurement settings.  To make this savings concrete we consider the number of Pauli terms one would need to measure for a 12-qubit system.  If no grouping is applied then there are 276 measurement circuits.  With greedy grouping considering locally commuting Pauli terms then there are 149 different measurement circuits.  With the measurement strategy outlined above we require 13 different circuits.}
\subsection{Off-diagonal terms with post-selection}
The circuits depicted in \figs{k1_opdm} did not allow for post-selection because the rotations to measure in the $X$-basis and $Y$-basis do not commute with the total number operator.  In this section we design a basis rotation circuit that commutes with the total number operator and diagonalizes the $\frac{1}{2}\left(XX + YY\right)$ Hamiltonian.  The diagonal form means that after performing the basis rotation we can measure in the computational basis to obtain expectation values $\frac{1}{2}\langle XX + YY\rangle$.  

\begin{figure}
    \centering
    \includegraphics[width=3cm]{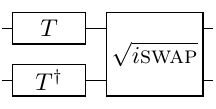}
    \caption{Two-mode fermionic fast Fourier transform that diagonalizes the $XX + YY$ Hamiltonian.}
    \label{fig:xy_measurement}
\end{figure}

\begin{figure}[H]
    \centering
    \includegraphics[width=12cm]{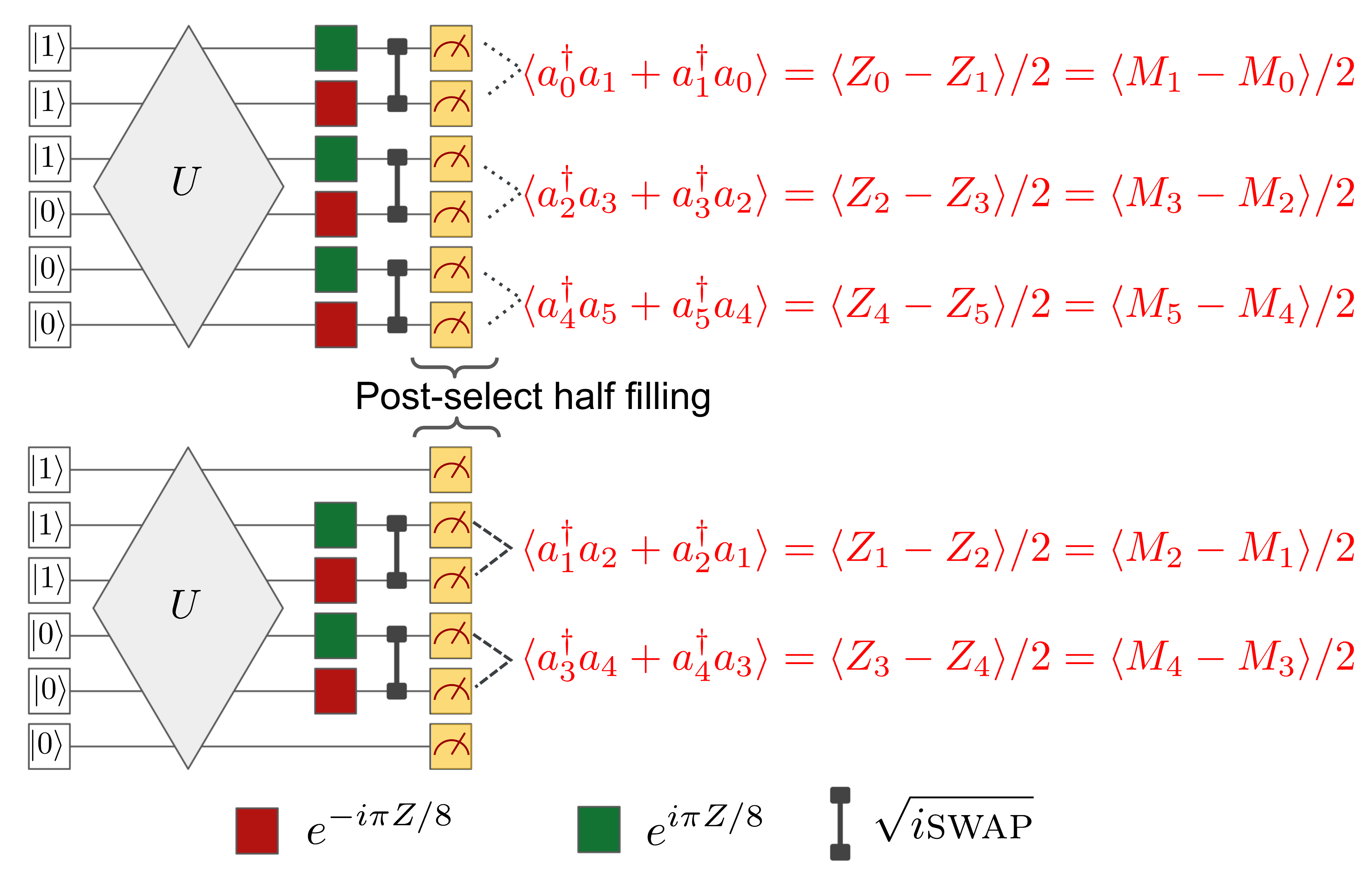}
    \caption{Two circuit measuring the one-off-diagonal of the $1$-RDM such that the total particle number can be measured simultaneously.  This \nick{circuit} allows us to post select on the correct number of excitations in the measured bitstring.  The top circuit measures the even pairs \nick{and} the bottom circuit measures the odd pairs.  Local $Z$ expectation values are measured on all the qubits and used to construct the expecation value for $\langle a_{i}^{\dagger}a_{i+1}\rangle$.}
    \label{fig:particle_conserving_circuit}
\end{figure}

The circuit that diagonalizes $\frac{1}{2}\left(XX + YY\right)$ is described in \figs{xy_measurement} and is denoted $U_{M}$ below.  Its commutation with the total number operator can be easy seen by recognizing that the $T$-gate (Rz$(\pi/4)$) commutes with the total number operator and so does the $\sqrt{i\textsc{swap}}$.  Applying $U_{M}$ to the $\frac{1}{2}\left(XX + YY\right)$ Hamiltonian 
\begin{align}\label{eq:measurement_similarity_transform}
U_{M} \begin{pmatrix}
0 & 0 & 0 & 0 \\
0 & 0 & 1 & 0 \\
0 & 1 & 0 & 0 \\
0 & 0 & 0 & 0
\end{pmatrix}
U_{M}^{\dagger} = 
\begin{pmatrix}
0 & 0 & 0 & 0 \\
0 & 1 & 0 & 0 \\
0 & 0 & -1 & 0 \\
0 & 0 & 0 & 0
\end{pmatrix}
\end{align}
transforms the operator into a diagonal representation. Given an ordered pair of qubits $\{a, a+1\}$ the last matrix in \eqref{eq:measurement_similarity_transform} is $\frac{1}{2}\left(Z_{a} - Z_{a+1}\right)$ in qubit representation.  Finally, we can relate the $Z$ expectation values, the transformed $XX + YY$ expectation values, fermionic ladder operators, and binary measurements $\{M_{a}, M_{a+1}\}$ via
\begin{align}
\langle U_{m} \left(a_{a}^{\dagger}a_{a+1} + a_{a+1}^{\dagger}a_{a}\right) U_{m}^{\dagger} \rangle = \langle U_{m} \frac{1}{2}\left(X_{a}X_{a+1} + Y_{a}Y_{a+1} \right) U_{m}^{\dagger} \rangle = \frac{1}{2}\langle Z_{a} - Z_{a+1} \rangle = \frac{1}{2}\left(M_{a+1} - M_{a}\right).
\end{align}
The measurement circuit can only be applied to non-overlapping pairs and thus we can obtain estimates of $X_{a}X_{a+1} + Y_{a}Y_{a+1}$ for $a$ values corresponding to even integers or $a$ corresponding to odd integers.  More concretely, we describe this process in \figs{particle_conserving_circuit} for a six qubit problem. \nick{All experiments involved circuits that allowed for post-selection based on total Hamming weight.  The ``raw'' data indicates analysis of the resulting bitstrings without post-selection.}
\subsection{Computing error bars for elements of the $1$-RDM}\label{app:error_bar_computation}
We use two methods to estimate error bars for all quantities in our experiments.  The procedures differ in how the covariance between $1$-RDM terms is estimated. In the first procedure, error bars are generated by estimating the covariance between terms in the $1$-RDM at the same time as the mean estimation. 
Mean values of off-diagonal $1$-RDM terms involve estimating the expectation values for $(Z_{a} - Z_{b})/2$. Therefore, the 
covariance between two off-diagonal elements of the $1$-RDM is 
\begin{align}
\mathrm{Cov}\left[\frac{1}{2}\left(Z_{a} - Z_{b}\right), \frac{1}{2}\left(Z_{p} - Z_{q}\right)\right] = \frac{1}{4}\left( \mathrm{Cov}\left[Z_{a}, Z_{p}\right] - \mathrm{Cov}\left[Z_{a}, Z_{q}\right] - \mathrm{Cov}\left[Z_{b},Z_{p}\right] + \mathrm{Cov}\left[Z_{b},Z_{q}\right]\right)
\end{align}
for all pair sets $\{(a, b), (p, q)\}$ measured simultaneously. All quantities can be estimated from the simultaneous measurement of all qubits.  Therefore, for each circuit permutation we obtain two covariance matrix of size $N/2 \times N/2$ and $N/2 - 1 \times N/2 -1$.  For the circuit with no label permutation we also obtain the \nick{covariances} for all $a_{i}^{\dagger}a_{i}$ terms.

In the second procedure for estimating covariance matrices we assume we are sampling from a pure Gaussian state. This assumption is applicable when the fidelity is high enough as any change to the covariance matrix would be a second order effect.  For these states the $2$-RDM is exactly described by the $1$-RDM and therefore all covariances between the $1$-RDM elements are perfectly defined by a non-linear function of the $1$-RDM elements.  For any wavefunction $\psi$ corresponding to the output of a basis rotation circuit the covariance of $1$-RDM elements computed from such a wavefunction are as follows:
\begin{align}
\mathrm{Cov}\left[a_{i}^{\dagger}a_{j} + a_{j}^{\dagger}a_{i}, a_{p}^{\dagger}a_{q} + a_{q}^{\dagger}a_{p}\right]_{\psi} = D_{q}^{i}\delta_{p}^{j} - D_{q}^{i}D_{j}^{p} + D_{p}^{i}\delta_{q}^{j} - D_{p}^{i}D_{j}^{q} + D_{q}^{j}\delta_{p}^{i} - D_{q}^{j}D_{i}^{p} + D_{p}^{j}\delta_{q}^{i} - D_{p}^{j}D_{i}^{q}.
\end{align}

With the estimates of the covariances we are able to re-sample the $1$-RDM assuming central-limit theorem statistics. We use a multinomial distribution where the mean values are $\langle a_{\sigma(i)}^{\dagger}a_{\sigma(i+1)}\rangle$ and the covariance matrix of the multinomial distribution is obtained by dividing the estimates of the covariance matrix above by $\alpha\times$ 250,000.  $\alpha$ is a number less than 1 reflecting the probability that a bitstring is rejected. $\alpha$ is estimated from  prior $N$-qubit experiments.  Once the new $1$-RDM is obtained it can be purified, used to estimate a fidelity witness, and compute the energy.  For all error bars we re-sample the $1$-RDM 1000 times and compute a mean value and standard deviation from this set.  All quantities estimated are sensitive to the $N$-representability of the resampled $1$-RDM. We use the fixed trace positive projection described in \cite{rubin2018application} to ensure that each resampled $1$-RDM is positive semidefinite and has the correct trace.  The correction procedure is only applied when the resampled $1$-RDM has eigenvalues below zero.

\section{Computing the fidelity and a fidelity witness from the $1$-RDM}\label{app:fidelity_fidelity_witness}
\subsection{Fidelity Witness}
The class of quantum circuits simulating \nick{non-interacting} fermion dynamics have the special property that an efficient fidelity witness can be derived.  The formal derivation for general \nick{non-interacting} fermion wavefunctions is described in Ref.~\cite{PhysRevLett.120.190501}.  Here we adapt this result to the special case of particle conserving dynamics generated by one-body fermionic generators.  A fidelity witness is an observable that provides a strict lower bound to the fidelity for all input states.  The fidelity witness is efficient in the sense that for an $L$-qubit system only $L^{2}$ expectation values are required to evaluate the fidelity witness.   Given that $U$ is a unitary corresponding to a basis transformation circuit and $\vert \omega \rangle$ is the initial computational basis state corresponding to $\mathbf{\omega} = (\omega_{1}, ..., \omega_{L})$ any $L$-bit string which satisfies $n_{j}\vert \mathbf{\omega} \rangle = \omega_{j}\vert \mathbf{\omega} \rangle$ for $j = 1, ..., L$ allows us to define a basis state annihilator operator
\begin{align}
n^{(\mathbf{\omega})} = \sum_{j=1}^{L}\left[ (1 - \omega_{j})n_{j} + \omega_{j}(\mathbb{I} - n_{j})\right] 
= \sum_{j=1}^{L}\left[n_{j} - \omega_{j}n_{j} + \omega_{j}\mathbb{I} - \omega_{j}n_{j}\right]
= \sum_{j=1}^{L} \left[n_{j} + \omega_{j}\mathbb{I} - 2 \omega_{j}n_{j}\right]
\end{align}
which satisfies $n^{(\mathbf{\omega})}\vert \mathbf{\omega} \rangle = 0$.  The computational basis state $\vert \mathbf{\omega} \rangle$ is the zero energy eigenstate of $n^{\mathbf{\omega}}$ and any other computational basis state an excitation from this state.  The excitation energy is exactly the number of bits that are different from $\omega$ for each Fock basis state which can be computed by summing the resulting bit string from the XOR operation between the two Fock basis states being considered.
The fidelty witness
\begin{align}
\mathcal{W} = U\left(\mathbb{I} - n^{\mathbf{\omega}}\right)U^{\dagger}
\end{align}
can be evaluated with knowledge of the measured $1$-RDM.  To relate the fidelity witness to the $1$-RDM it is important to note the following
\begin{align}
\mathrm{Tr}\left[ U \rho_{p} U^{\dagger} a_{i}^{\dagger}a_{j}\right] = \left[ \mathbf{u} D \mathbf{u}^{\dagger}\right]_{i, j}
\end{align}
where $D$ is the matrix of expectation values $\langle \rho_{p}, a_{i}^{\dagger}a_{j}\rangle$ and $\mathbf{u} = e^{\kappa}$ because any one-body rotation on the state $\rho_{p}$ can be equated to a similarity transform of the generating matrix for that one-body transformation. \nick{This logic is similar to} logic used in~\cite{PhysRevX.10.011004} \nick{which} move\nick{d} one-body basis rotations at the end of the circuit into the Hamiltonian as an error mitigation technique. Using this relationship we can evaluate the fidelity witness with the following expression
\begin{align}
F_{\mathcal{W}}(\rho_{p}) = 1 - \sum_{j=1}^{L} \left(\left[ \mathbf{u}^{\dagger} \mathbf{D} \mathbf{u}\right]_{j, j} + \omega_{j} - 2 \omega_{j} \left[ \mathbf{u}^{\dagger} \mathbf{D} \mathbf{u}\right]_{j, j} \right)
\end{align}
where $\mathbf{D}$ is the 1-RDM that is measured, $\mathbf{u} = e^{\mathbf{\kappa}}$ is the unitary rotation representing the new Slater determinant.
\subsection{Fidelity}\label{app:fidelity_measure}
Given an idempotent $1$-RDM and the basis rotation unitary $u = e^{\kappa}$, the fidelity can be determined by the following procedure:
\begin{enumerate}
    \item[1.] Perform an eigen decomposition on the purified $1$-RDM and use the eigenvectors associated with eigenvalues equal to $1$ as the columns of a unitary matrix $v$ corresponding to the measured basis rotation.
    \item[2.] Use the expression for the overlap between two basis rotation unitaries $\vert \langle \psi_{u} \vert \psi_{v} \rangle \vert^{2} = \vert \mathrm{det}\left(v^{\dagger}u\right) \vert^{2}$ to compute the fidelity. The function $\mathrm{det}$ is the determinant of a matrix.  This is the inner product between two Grassmann representatives and is independent of choice of orbitals.
\end{enumerate}
\section{Error mitigation through purification}\label{app:purification}
A \nick{distinctive} feature of the Slater determinant wavefunction ansatz is that their $1$-RDMs are idempotent matrices.  The manifold of states with idempotent $1$-particle density matrices is significantly smaller than the space of possible wavefunctions.  Thus our error mitigation strategy will rely on projecting the measured $1$-RDM to the closest idempotent $1$-RDM. This projection procedure can be represented by the following mathematical program
\begin{align}\label{eq:rdm_projection}
\min_{Tr[D] = \eta, D \succeq 0, D^{2} = D} ||D - \tilde{D}||
\end{align}
that seeks to determine a $1$-RDM $D$ that is close to the measured $1$-RDM $\tilde{D}$ with has fixed trace, is positive semidefinite, and is a projector.  A practical implementation of the the program in \eq{rdm_projection} is challenging due to the idempotency constraint.  Instead of solving \eq{rdm_projection} directly we rely on an iterative procedure that under mild conditions projects a measured $1$-RDM $\tilde{D}$ towards the set of idempotent matrices.  This procedure is the McWeeny purification commonly used in linear scaling electronic structure techniques~\cite{RevModPhys.32.335} and is defined by the iteration
\begin{align}\label{eq:mcweeny_process}
D_{n+1} = 3 D^{2}_{n} - 2 D^{3}_{n}.
\end{align}
After each iteration the eigenvalues are closer to $\{0, 1\}$.  Prior work~\cite{mccaskey2019quantum} proposed to use McWeeny purification on the $2$-RDM, but it is not clear what that accomplishes. This is because, in general, $2$-RDMs are not idempotent matrices and applying pure-state purification requires more general pure-state $N$-representability conditions~\cite{klyachko2006quantum}. \nick{Due to the fact that McWeeny iteration has no effect on the eigenvectors, it merely pushes the eigenvalues of $D$ towards $\{0, 1\}$, we could have achieved this projection by diagonalizing $D$ and rounding the eigenvalues to $0$ or $1$. We performed the purification iteration because we are able to analyze the convergence when $D$ is obtained by sampling.}


Here we will estimate the number of samples needed to ensure that the $1$-RDMs can be faithfully reconstructed within arbitrarily small error using our protocol.  This analysis assumes we sample from a perfect state and thus our goal is to provide evidence that McWeeny purification is convergent under sampling noise.
Consider the purification process in \eq{mcweeny_process}.  Now let us assume that the principal eigenvalue of $D$ is $P_k$.  In absentia of numerical error we would have that $P_k=1$ for Hartree-Fock theory.  However, sampling error incurs an error in this eigenvalue such that
\begin{equation}
    P_k = 1 + \Delta,
\end{equation}
where $\Delta$ is a random variable with mean $0$ and variance $\sigma^2$.  Further, let $\mu_k = \mathbb{E} (\Delta^k)$, where $\mu_2 = \sigma^2$ for example.  Now given these quantities we wish to evaluate
\begin{equation}
    \mathbb{E} (P_{k+1}) = \mathbb{E}(3P_k^2 -2 P_k^3) = 1-3\sigma^2 -2\mu^3.
\end{equation}
Similarly we have that
\begin{equation}
    \mathbb{E}(P_{k+1}^2) = 1-6\sigma^2 -4\mu_3 +9\mu_4 +12\mu_5 +4\mu_6.
\end{equation}
This implies that the variance is
\begin{equation}
    \mathbb{V}(P_{k+1}) = \mathbb{E}(P_{k+1}^2) - \mathbb{E}(P_{k + 1})^2 = 9(\mu_4 - \sigma^4) +12(\mu_5 - \sigma^2\mu_3) +4(\mu_6 - \mu_3^2).
\end{equation}
Further, let us assume that $\mu_j \le \alpha_j \sigma^j$, for all $j$ .
\begin{equation}
    \mathbb{V}(P_{k+1}) \le 9 \sigma^4 (\alpha_4-1) + 12|\sigma|^5 (\alpha_5 + \alpha_3)+4 |\sigma|^6\alpha_6.
\end{equation}
Assuming that $\sigma \le 1$ we have that
\begin{equation}
    \mathbb{V}(P_{k+1}) \le 9 \sigma^4 (\alpha_4-1) + 4|\sigma|^5 (\alpha_6+3\alpha_5 + 3\alpha_3).
\end{equation}
It is clear from this recurrence relation that the variance for this method converges quadratically (assuming $\sigma$ is sufficiently small).  Specifically, we have that $\mathbb{V}(P_K)\le \epsilon$ for $K\in \mathcal{O}\left(\log\log(1/\epsilon) \right)$ if appropriate convergence criteria are met.
A criterion for convergence is that $9 \sigma^4 (\alpha_4-1) + 4|\sigma|^5 (\alpha_6+3\alpha_5 + 3\alpha_3)\le \sigma^2$.  This is guaranteed if,
\begin{equation}
    \sigma^2 \le \frac{1}{9(\alpha_4-1)}\left(1 - \frac{\beta\left(-\beta + \sqrt{\beta^2 +36\alpha_4 -36} \right)}{18(\alpha_4-1)} \right),\label{eq:non-gaussian}
\end{equation}
where $\beta = 4(\alpha_6+3\alpha_5+3\alpha_3)$.

The precise values of $\alpha_j$ depend on the nature of the underlying distribution.  However, if we assume that it is Gaussian then we have that $\alpha_{2j+1} =0~\forall~j$, $\alpha_4=3$, $\alpha_6=15$.  Furthermore, we have under these Gaussian assumptions (for any $\sigma>0$) that
\begin{equation}
    \mathbb{V}(P_{k+1}) \le 18 \sigma^4  + 12\sigma^6.\label{eq:sigma2gauss}
\end{equation}
In this case, we find that the McWeeny iteration will converge if $\mathbb{V}(P_{k+1})\le\sigma^2$ which is implied by
\begin{equation}
    \sigma^2 \le \frac{-3}{20}+\frac{\sqrt{564}}{120}\approx 0.048.
\end{equation}
This relatively broad distribution implies that even if the uncertainty in the principal eigenvalue of the reconstructed RDM is large then the algorithm will with high probability converge to a pure state after a small number of iterations (if the underlying distribution is Gaussian).  If the distribution is non-Gaussian then \eq{non-gaussian} can be used to show convergence given that the moments of the distribution are appropriately small.
\subsection{Errors in Eigenvalues}
The errors in the eigenvalues of the RDM are easy to compute from known results.  We have from Corollary 6.3.4 from~\cite{horn2012matrix} that if $\rho$ is the true density operator and $\tilde{\rho} = \rho + sE$ for some matrix $E$ of errors and some scalar $s\in [0,1]$ then the error in a particular eigenvalue is at most
\begin{equation}
    |\lambda(\rho) - \lambda(\rho + sE) | \le s\|E\|,\label{eq:evpert}
\end{equation}
where $\|E\|$ is the spectral norm of the error matrix.
We are of course most interested in the case where $s=1$, however below we will need the above formula for general values of $s$ and so we give it for generality.

Now let $E$ be a matrix consisting of $M$ elements, each of which is independently distributed with zero mean and variances at most $\sigma^2_M$.  We then have that 
\begin{equation}
    \mathbb{E}\left((\lambda(\rho) - \lambda(\rho + E))^2\right)\le \mathbb{E}\left(\sum_{i,j} [E^2]_{i,j} \right)\le M \sigma^2_M.
\end{equation}
Thus
\begin{equation}
    \mathbb{V}(\lambda(\rho +E)) \le M \sigma_M^2.\label{eq:varlambda}
\end{equation}
Hence $\sigma^2 \le M\sigma_M^2$, which allows the upper bounds in \eq{sigma2gauss} to be easily computed (under assumptions of Gaussianity).  In particular, we then have convergence under the Gaussianity assumption if
\begin{equation}
    \sigma_M^2 \le \frac{1}{M}\left(\frac{-3}{20} + \frac{\sqrt{564}}{120} \right).
\end{equation}
Recall that the $1$-RDM constists of $N(N+1)/2$ independent matrix elements, which implies that $M=N(N+1)/2$ in our case.
\subsection{Errors in Eigenvectors}
\nick{Although} the above criteria give conditions for the convergence of McWeeny purification starting from a sampled $1$-RDM, there remains the question of whether the pure state that it converges to is $\epsilon$-close to the true value.  This is relevant because if the errors are large enough that an eigenvalue crossing occurs, then the purification process can fail to yield the desired state.  Our aim here is to bound the distance between the eigenvectors.

First, rather than arguing about the difference in eigenvectors for $\rho$ and $\rho+E$ we will instead consider $R$ time slices and will be interested in the eigenvectors of $\rho(j) :=\rho + (j/R)E$.  Let the principal eigenvector of $\rho$ be $\ket{\lambda}$ and more generally at step $j$ let us denote the eigenvector to be $\ket{\lambda(j)}$ and the correspeonding eigenvalue to be $\lambda(j)$.  We then have from first order perturbation theory, assuming that there is an eigenvalue gap that for any state $\ket{\nu([j-1])}$ that is orthogonal to $\ket{\lambda([j-1])}$,
\begin{equation}
    \brakket{\nu(j-1)}{\lambda(j)} = \frac{1}{R} \frac{\bra{\nu(j-1)} E \ket{\lambda(j-1)}}{\nu(j-1) - \lambda(j-1)}+ O(1/R^2)
\end{equation}
Thus if we define $\gamma(j)$ to be the minimum eigenvalue gap between $\ket{\lambda(j)}$ and the remainder of the spectrum of $\rho(j)$ we have that
\begin{align}
    \sum_{\nu \ne \lambda} |\brakket{\nu(j-1)}{\lambda(j)}|^2 &\le \sum_{\nu \ne \lambda} \frac{1}{R^2} \frac{|\bra{\nu(j-1)} E \ket{\lambda(j-1)}|^2}{(\lambda(j-1) - \nu(j-1))^2}+ O(1/R^3)\nonumber\\
    &\le \sum_{\nu \ne \lambda} \frac{1}{\gamma^2(j-1)R^2} |\bra{\nu(j-1)} E \ket{\lambda(j-1)}|^2+ O(1/R^3)\nonumber\\
    &=\sum_{\nu \ne \lambda} \frac{1}{\gamma^2(j-1)R^2} \bra{\lambda(j-1)}E\ket{\nu(j-1)}\!\!\bra{\nu(j-1)} E \ket{\lambda(j-1)}|^2+ O(1/R^3)\nonumber\\
    &= \frac{1}{\gamma^2(j-1)R^2} \left(\bra{\lambda(j-1)} E^2 \ket{\lambda(j-1)} - (\bra{\lambda(j-1)} E \ket{\lambda(j-1)})^2\right) + O(1/R^3)\nonumber\\
    &\le \frac{\|E^2\|}{\gamma^2(j-1)R^2} + O(1/R^3)\label{eq:sumcomp}
\end{align}
It then follows from \eq{sumcomp} that
\begin{equation}
    |\brakket{\lambda(j-1)}{\lambda(j)}-1|^2 \le \frac{\|E^2\|}{\gamma^2(j-1)R^2} + O(1/R^3)
\end{equation}
This gives us that, for the Euclidean distance between two vectors,
\begin{equation}
    \left|\ket{\lambda(j)} - \ket{\lambda(j-1)} \right| \le \frac{\sqrt{2\|E^2\|} }{\gamma(j-1) R}+O(1/R^2).
\end{equation}
Next we have from the triangle inequality that for any integer $R$,
\begin{equation}
    \left|\ket{\lambda(R)} - \ket{\lambda(0)} \right| \le \sum_{j=1}^R \left|\ket{\lambda(j)} - \ket{\lambda(j-1)} \right| \le \sum_{j=1}^R\frac{\sqrt{2\|E^2\|} }{\gamma(j-1) R}+O(1/R)
\end{equation}
In particular, this holds as we take $R\rightarrow \infty$, which yields
\begin{equation}
     \lim_{R\rightarrow \infty}\sum_{j=1}^R\frac{\sqrt{2\|E^2\|} }{\gamma(j-1) R}+O(1/R) \le \frac{\sqrt{2\|E^2\|} }{\gamma_{\min}}=\frac{\sqrt{2}\|E\| }{\gamma_{\min}}.
\end{equation}
Unfortunately, we do not know what $\gamma_{\min}$ is apriori, however we can bound it modulo some weak assumptions.  Let $\|E\| \le 1/4$, it is then straight forward to verify from \eq{evpert} that
\begin{equation}
    \gamma_{\min}\ge 1-2\|E\|.
\end{equation}
Under the exact same assumptions we then have from a series expansion of the denominator that
\begin{equation}
    \left|\ket{\lambda(\rho)} - \ket{\lambda(\rho + E)} \right| \le {\sqrt{2} \|E\|}\left(1+4\|E\| \right) \le 2\sqrt{2} \|E\|
\end{equation}
As $E$ is a sum of $M$ elements each with zero mean and variance at most $\sigma_M$ we then have under the above assumptions (and the additive property of variance) that
\begin{equation}
    \mathbb{V}\left|\ket{\lambda(\rho)} - \ket{\lambda(\rho + E)} \right| \le 8 M \sigma_M^2.
\end{equation}
Therefore if we demand that the variance is atmost $\epsilon^2$ it suffices to pick
\begin{equation}
    \sigma_M^2 = \frac{\epsilon^2}{8M}, \label{eq:sigmachoice}
\end{equation}
which sets a sufficient condition on the number of samples of $N_{\rm samp} \ge \frac{\epsilon}{2\sqrt{2M}}$.  The remaining caveat is that in the above analysis we needed to assume that $\|E\|\le 1/4$.  If each of the entries of the matrix $E$ are Gaussian random variables, for example, it then follows that regardless of the value of $\sigma$ there will always be a tail probability that this eigenvalue condition is not met.  We can bound the tail probability using Chebyshev's inequality.  Using the exact same reasoning as in \eq{varlambda} we have that
\begin{equation}
    \mathbb{V}(\|E\|) \le M \sigma_M^2.
\end{equation}
Thus the probability that $\|E\| \ge 1/4$ is
\begin{equation}
    P \le {16M\sigma_M^2}\sim {2\epsilon^2}.
\end{equation}
Thus even under the pessimistic assumptions of Chebyshev's inequality, we have that the probability of failure is asymptotically negligible if~$\sigma_M$ is chosen in accordance with \eq{sigmachoice}.  Note that the number of samples needed taken in this case is in $\Theta(\epsilon/N)$ as there are $M\in \Theta(N^2)$ independent matrix elements in the $1$-RDM.

\section{Effect of $\textsc{CPHASE}$ and Givens Rotation Error}\label{app:parasitic_cphase_analysis}

In this section we consider two known gate errors that occur in the Givens rotation circuits and attempt to analytically and numerically benchmark the effect of these errors.  
When implementing the $\sqrt{i\textsc{swap}}$ operation there is a known $\vert 11 \rangle \langle 11\vert$ phase error of approximately $\pi/24$.  We model this phase error as a $\textsc{cphase}(\pi/24)$ gate that occurs directly after the $\sqrt{i\textsc{swap}}$ gate (Eq.~\eqref{eq:effective_iswap}).  We find that the always on $\textsc{cphase}(\pi/24)$ has negligible effect on the outcome of the experiment \nick{and} the stochastic $\mathrm{Rz}(\theta)$ errors coherently corrupt the output of the circuit. 
\begin{align}\label{eq:effective_iswap}
\sqrt{i\textsc{swap}} \approx
\begin{pmatrix}
1 & 0 & 0 & 0 \\
0 & \frac{1}{\sqrt{2}} & \frac{i}{\sqrt{2}} & 0 \\
0 & \frac{i}{\sqrt{2}} & \frac{1}{\sqrt{2}} & 0 \\
0 & 0 & 0 & e^{i \pi / 24} \\
\end{pmatrix} =
\textsc{CPHASE}(\pi/24)\sqrt{i\textsc{swap}}
\end{align}
To benchmark the effect of the parasitic $\textsc{cphase}$ we simulate the diazene experiment with this interaction turned on and evaluate the results with error mitigation.  
\begin{figure}[htb]
    \centering
    \includegraphics[width=8.5cm]{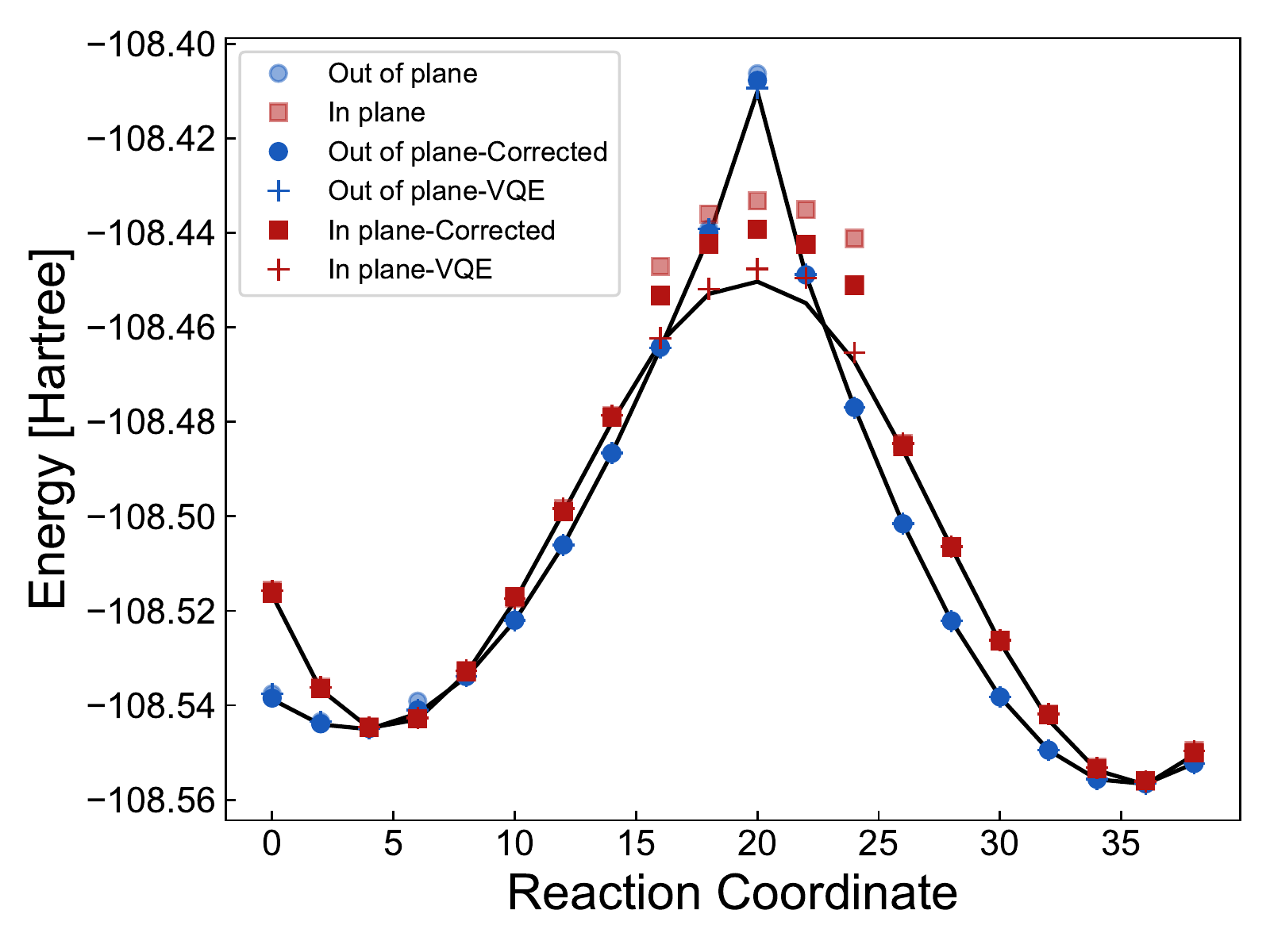}
    \caption{Analysis of the diazene isomerization curve where the Givens rotations are corrupted by a parasitic $\textsc{cphase}(\pi/24)$.  \nick{All points are after purification. Without purification all curves are significantly higher in energy.} The light dots are the circuits executed without optimization, the darker dots are with an angle adjustment to counteract the known parasitic $\textsc{cphase}$, and the plus markers are VQE optimization of the $\textsc{cphase}$ circuits.  }
    \label{fig:cphase_analysis_diazene}
\end{figure}
We can counteract the $\textsc{cphase}(\pi/24)$ by performing local Rz gates.  Consider the imperfect gate
\begin{align}
U_{1} = \mathrm{diag}(1, 1, 1, e^{-i \phi})\sqrt{i\textsc{swap}}
\end{align}
we can use a different imperfect gate which differs only by single qubit phases, 
\begin{align}
U_{2} = \mathrm{diag}(1, e^{i \phi/2}, e^{i\phi/2}, 1)\sqrt{i\textsc{swap}}.
\end{align}
The error associated with $U_{1}$ can be approximated by considering the Pauli expansion of the $\textsc{cphase}$ part of $U_{1}$,
\begin{align}
\mathrm{diag}(1, 1, 1, e^{-i \phi}) \approx e^{-i\phi} \times \left(II + \frac{i\phi}{4} IZ + \frac{i\phi}{4}ZI - \frac{i\phi}{4}ZZ\right),
\end{align}
and thus the error is approximately
\begin{align}
\mathrm{Err}_{1} \approx 3 \left(\frac{\phi}{4}\right)^{2} = \frac{3}{16}\phi^{2}.
\end{align}
Similarly for $U_{2}$
\begin{align}
\mathrm{diag}(1, e^{i \phi/2}, e^{i\phi/2}, 1) \approx e^{i\phi/4}\left(II - \frac{i\phi}{4}ZZ\right)
\end{align}
with an associated Pauli error of
\begin{align}
\mathrm{Err}_{2} \approx \frac{1}{16}\phi^{2}.
\end{align}
Very crudely, since for $\sqrt{i\textsc{swap}}$ $\phi=\pi/24$, we expect $\mathrm{Err}_{1}$ to be approximately $0.32\%$ and $\mathrm{Err}_{2}$ to be approximately $0.11\%$.  This improvement is shown to be most beneficial for simulating the in plane rotations of diazene in ~\figs{cphase_analysis_diazene}.  The light dots are with the original $\textsc{cphase}$ gate whereas the solid dots are with this local $\mathrm{Rz}$ correction.  We also include a VQE optimization to numerically determine the noise floor for this experiment.  This suggests that VQE + error mitigation can mitigate not only control error but more fundamental gate physics issues. For in-plane rotation circuits the dynamics during the circuit execution are apparently more sensitive to these types of coherent errors near transition states, although the exact reason for increased sensitivity is unclear.

To determine the error budget on the Rz rotation angles we can determine the degree of corruption from Gaussian noise on the control angle.  Consider the Rz rotation 
\begin{align}
\mathrm{Rz}(\theta, \delta\alpha) = e^{-i Z \theta (1 + \delta\alpha) / 2}
\end{align}
where $\theta$ is the desired rotation angle and $\delta\alpha$ is a stochastic variable.  We can build a simplified model of control angle error as Givens rotation error
\begin{align}\label{eq:givens_control_error}
G(\theta, \delta\alpha) = e^{\theta (1 + \delta\alpha) \left(a_{i}^{\dagger}a_{j} - a_{j}^{\dagger}a_{i}\right)}
\end{align}
which can be expressed as 
\begin{align}
G(\theta) = \sqrt{i\textsc{swap}}_{i,j}^{\dagger}e^{-i\theta ( 1 + \delta\alpha) Z_{i} /2}e^{i\theta( 1 + \delta\alpha) Z_{j} /2}\sqrt{i\textsc{swap}}_{i,j}.
\end{align}
For numerical simplicity we consider the effect on elements of the $1$-RDM

\begin{align}
G(-\theta, \delta\alpha, i, j)a_{r}^{\dagger}G(\theta, \delta\alpha, i, j) = 
\begin{cases}
a_{i}^{\dagger}\cos(\theta (1 + \delta\alpha)) + a_{j}^{\dagger}\sin(\theta (1 + \delta\alpha)) & \mathrm{if}\;r = i\\
a_{j}^{\dagger}\cos(\theta (1 + \delta\alpha)) - a_{i}^{\dagger}\sin(\theta (1 + \delta\alpha)) & \mathrm{if}\;r = j \\
a_{r}^{\dagger} & \mathrm{if}\;r \neq i \And r \neq j
\end{cases}
\end{align}
\begin{align}
G(-\theta, \delta\alpha, i, j)a_{s}G(\theta, \delta\alpha, i, j) = 
\begin{cases}
a_{i}\cos(\theta (1 + \delta\alpha)) + a_{j}\sin(\theta(1 + \delta\alpha)) & \mathrm{if}\;s = i\\
a_{j}\cos(\theta (1 + \delta\alpha)) - a_{i}\sin(\theta(1 + \delta\alpha)) & \mathrm{if}\;s = j \\
a_{s} & \mathrm{if}\;s \neq i \And s \neq j
\end{cases}
\end{align}
We can determine the expected $1$-RDM with respect to a Gaussian distribution of noise by integrating with respect to the perturbation
\begin{align}
\rho(\delta\alpha) = \frac{1}{\sigma\sqrt{2\pi}} e^{-\frac{\left(\delta\alpha\right)^{2}}{2\sigma^{2}}}
\end{align}
\begin{align}
\int_{-\infty}^{\infty} \rho(\delta\alpha, \sigma) G(-\theta, \delta\alpha, i, j)a_{r}^{\dagger}G(\theta, \delta\alpha, i, j)d\delta\alpha = 
\begin{cases}
a_{i}^{\dagger}\cos(\theta)e^{-\frac{\theta^{2}\sigma^{2}}{2}} + a_{j}^{\dagger}\sin(\theta)e^{-\frac{\theta^{2}\sigma^{2}}{2}} & \mathrm{if}\;r = i\\
a_{j}^{\dagger}\cos(\theta)e^{-\frac{\theta^{2}\sigma^{2}}{2}} - a_{i}^{\dagger}\sin(\theta)e^{-\frac{\theta^{2}\sigma^{2}}{2}} & \mathrm{if}\;r = j \\
a_{r}^{\dagger} & \mathrm{if}\;r \neq i \And r \neq j
\end{cases}
\end{align}

\begin{align}\label{eq:noisy_givens_rotation}
\int_{-\infty}^{\infty} \rho(\delta\alpha, \sigma)G(-\theta, \delta\alpha, i, j)a_{s}G(\theta, \delta\alpha, i, j)d\delta\alpha = 
\begin{cases}
a_{i}\cos(\theta) e^{-\frac{\theta^{2}\sigma^{2}}{2}} + a_{j}\sin(\theta)e^{-\frac{\theta^{2}\sigma^{2}}{2}}  & \mathrm{if}\;s = i\\
a_{j}\cos(\theta)e^{-\frac{\theta^{2}\sigma^{2}}{2}}  - a_{i}\sin(\theta) e^{-\frac{\theta^{2}\sigma^{2}}{2}} & \mathrm{if}\;s = j \\
a_{s} & \mathrm{if}\;s \neq i \And s \neq j
\end{cases}
\end{align}
Therefore, propagating the $1$-RDM with stochastic Rz errors corresponds to evaluating the map in \eq{noisy_givens_rotation}.
\begin{figure}[htb]
    \centering
    \includegraphics[width=8.5cm]{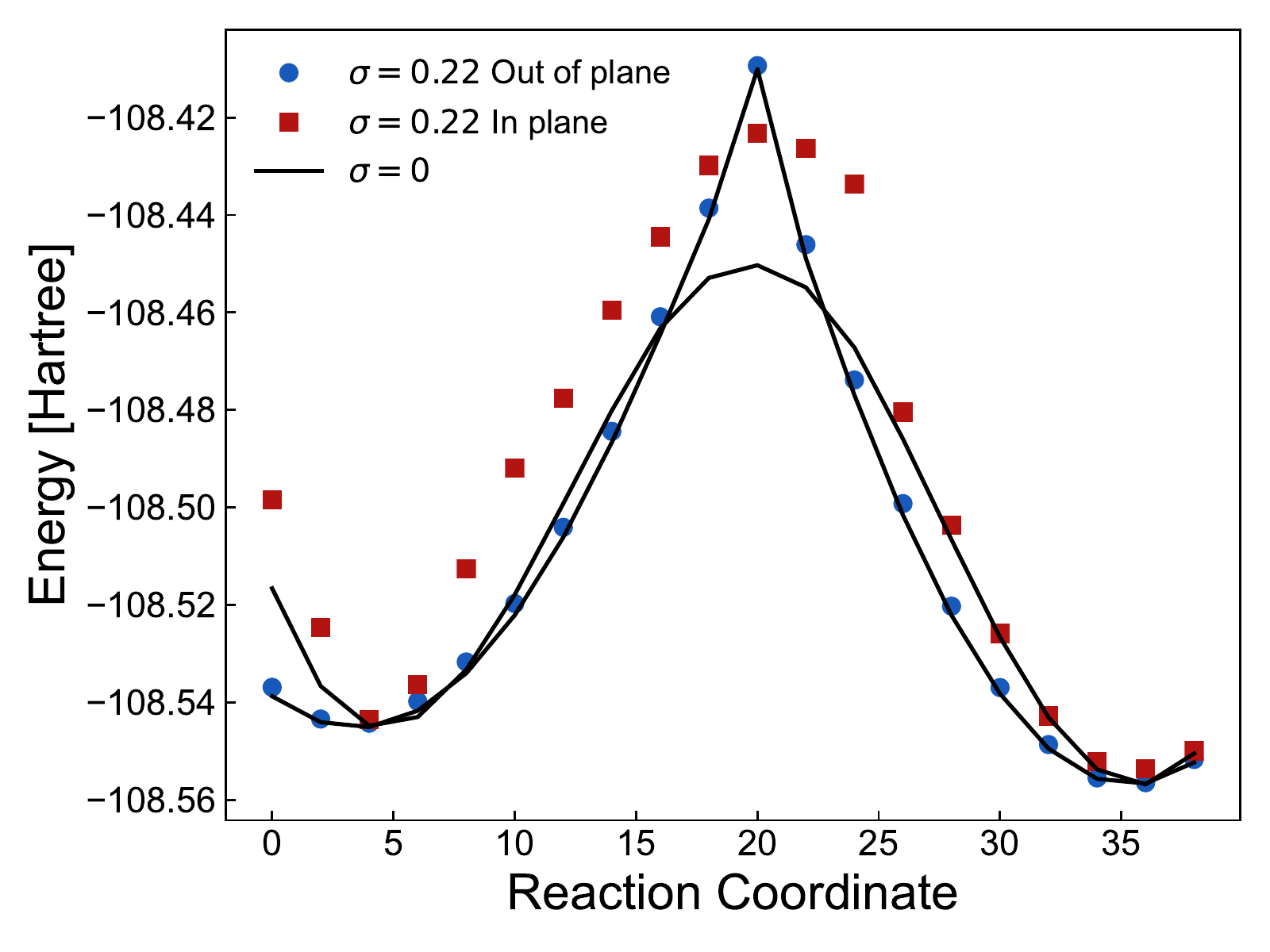}
    \caption{Stochastic $\mathrm{Rz}$ simulation of diazene with $\sigma = 0.22$ radian fluctuation on the Givens rotation gates. The plotted points are after applying purification to the result $1$-RDM}
    \label{fig:rz_error_analysis_diazene}
\end{figure}
This calculation assumes that the stochasticity has a time-scale that is much faster than a single energy evaluation.  We find that with $\sigma > 0.22$ purification projects to the wrong $1$-RDM.

\section{Gradient for the Basis Rotation Ansatz}\label{sec:grads_hf}
Another benefit of restricting our ansatz to Slater determinants is the fact that the gradient with respect to the parameters is accessible via the elements of the $1$-RDM. 
The gradient of the energy with respect to the parameters of a one-body generator $Z = \sum_{i < b}c_{b, i}\left(a_{b}^{\dagger}a_{i} - a_{i}^{\dagger}a_{b}\right)$ is
\begin{align}
\frac{d E}{d c_{b, i}} = \langle \phi_{0}\vert \frac{d e^{-Z}}{d c_{b, i}} H e^{Z} \vert \phi_{0} \rangle + \langle \phi_{0}\vert  e^{-Z} H \frac{d e^{Z}}{d c_{b, i}} \vert \phi_{0} \rangle.
\end{align}
Due to the structure of this operator we expect the gradient to involve the commutator of the Hamiltonian with respect to the anithermitian operator that becomes the prefactor to the right gradient.  We call this prefactor $\nabla f(Z)$ to indicate that it is a different operator from just the rotation generator associated with $c_{b, i}$.
\begin{align}
\frac{d E}{d c_{b, i}} = \langle \phi_{0}\vert e^{-Z} \left[H, \nabla f(Z)_{c_{b,i}}\right] e^{Z} \vert \phi_{0} \rangle 
\end{align}
All quantities in the commutator above can be evaluated with knowledge of the $1$-RDM when $\phi_{0}$ is a computational basis state.  In this work we utilized this gradient for a classical implementation and provide it here as justification for the ansatz and for future studies.  The formal derivation of $\nabla f(Z)$ can be found in \cite{wilcox1967exponential} and \cite{Helgaker2002}.  

As a sketch for the form of $\nabla f(Z)$ consider the unitary performed in the Hartree-Fock experiment
\begin{align}
U(c_{b, i}) = e^{\sum_{b, i}c_{b, i}E_{b, i}^{-}} = e^{\sum_{b, i}c_{b, i}(a_{b}^{\dagger}a_{i} - a_{i}^{\dagger}a_{b})}.
\end{align}
We now want to consider the energy derivative with respect to $c_{b,i}$.  Using the formulas in \cite{wilcox1967exponential} we obtain
\begin{align}\label{eq:uderiv}
\frac{d U(\mathbf{c})}{d c_{b, i}} = \left[\int_{0}^{1}dx e^{x\sum_{b, i}c_{b,i}E_{b, i}^{-}}E_{b, i}^{-}e^{-x\sum_{b, i}c_{b,i}E_{b, i}^{-}}\right]e^{\sum_{b, i}c_{b, i}E_{b, i}^{-}}.
\end{align}
In order to evaluate this integral we need to have an analytical form for the similarity transform of the integrand.  The integrand can be expressed in series form with the Baker-Campbell-Hausdorff identity where each term involves nested commutators.  Each nested commutator can be expressed more succinctly as the adjoint action of $Z$ on $E_{b,i}^{-}$
\begin{align}
\mathrm{ad}\left(\sum_{b', i'}c_{b', i'}E_{b', i'}^{-}\right)^{n}\left(E_{b, i}^{-}\right).
\end{align}
A general strategy for evaluating sums of adjoint actions is to represent the operator $\sum_{b', i'}c_{b', i'}E_{b', i'}^{-}$ in its eigenbasis and directly evaluate the commutator as a matrix power.   In our case this would involve diagonalizing a large $2^{n} \times 2^{n}$ matrix.  Fortunately, due to the connection between one-particle-basis rotations and rotations by one-body operators on the full Hilbert space we can find a $n \times n$ unitary that can diagonalize the matrix of $c_{b, i}$ coefficeints and represent the operator $E_{b, i}^{-}$ in this one-particle basis.  Following this step of the derivation in \cite{Helgaker2002} we form the $\mathbf{C}$ matrix of coefficients $c_{b, i}$ which is antihermitian and diagonalize.  Therefore, $\mathbf{C}$ is represented in its eigenbasis as
\begin{align}
i\mathbf{C} = i\sum_{r}\lambda_{r}\tilde{a}_{r}^{\dagger}\tilde{a}_{r}
\end{align}
where $\lambda$ are purely imaginary and we have used the fact that 
\begin{align}
\tilde{a}_{p} = \sum_{q}u_{p,q}^{*}a_{q} \;\;,\;\;\tilde{a}_{p}^{\dagger} = \sum_{q}u_{p,q}a_{q}^{\dagger}.
\end{align}
We represent $E_{b, i}^{-}$ term in the basis that diagonalizes $i\mathbf{C}$
\begin{align}
\mathbf{Y} = \sum_{k,l}Y_{kl}\tilde{a}_{k}^{\dagger}\tilde{a}_{l} \\
Y_{kl} = \left(U^{\dagger} E_{b, i}^{-} U\right)_{k,l}
\end{align}
here $E_{b, i}^{-}$ is an antisymmetric matrix with $1$ at the $(b, i)$ position and $-1$ at $(i, b)$ position which is a representation of the operator $E_{b, i}^{-}$. Therefore,
\begin{align}
\mathrm{ad}\left(\sum_{b', i'}c_{b', i'}E_{b', i'}^{-}\right)\left(E_{b, i}^{-}\right) =& \sum_{rkl}i \lambda_{r} Y_{kl}\left[\tilde{a}_{r}^{\dagger}\tilde{a}_{r},  \tilde{a}_{k}^{\dagger}\tilde{a}_{l}\right] \\
=& \sum_{rkl}i \lambda_{r} Y_{kl}\left(\tilde{a}_{r}^{\dagger}\tilde{a}_{l}\delta_{k}^{r} - \tilde{a}_{k}^{\dagger}\tilde{a}_{r}\delta_{r}^{l}\right)\\
=& i\sum_{kl} \left(\lambda_{k} - \lambda_{l} \right) Y_{kl}\tilde{a}_{k}^{\dagger}\tilde{a}_{l}.
\end{align}
Furthermore, powers of the adjoint action are
\begin{align}
\mathrm{ad}\left(\sum_{b', i'}c_{b', i'}E_{b', i'}^{-}\right)^{n}\left(E_{b, i}^{-}\right) =i^{n}\sum_{kl} \left(\lambda_{k} - \lambda_{l} \right)^{n} Y_{kl}\tilde{a}_{k}^{\dagger}\tilde{a}_{l}.
\end{align}
Armed with the adjoint power we can now evaluate the integrand of Eq.~\eqref{eq:uderiv} via fundemental theorem of calculus and arrive at an expression for the gradient
\begin{align}\label{eq:fermion_one_hop}
\frac{d U(\mathbf{c})}{d c_{b, i}} =& \left[\sum_{kl}Y_{kl}\frac{e^{i(\lambda_{k} - \lambda_{l})} - 1}{i\left(\lambda_{k} - \lambda_{l}\right)} \tilde{a}_{k}^{\dagger}\tilde{a}_{l}\right]e^{\sum_{b, i}c_{b, i}E_{b, i}^{-}} \\
=& \left(\sum_{k,l}\left[U M U^{\dagger}\right]_{kl} a_{k}^{\dagger}a_{l}\right) e^{\sum_{b, i}c_{b, i}E_{b, i}^{-}}
\end{align}
where $M_{kl} = Y_{kl} \frac{e^{i(\lambda_{k} - \lambda_{l})} - 1}{i\left(\lambda_{k} - \lambda_{l}\right)}$.  The expression in the parenthesis is a new one-body operator that we previous denoted $\nabla f(Z)$.

\section{Optimization Technique}\label{app:iterative_opt}
The optimizer we use in the experiment is based on Kutzelnigg's approach to iteratively constructing a wavefunction that satisfies the Brillouin condition~\cite{kutzelnigg1979generalized}.  In the following section we include the derivation and modifications of this procedure from Reference~\cite{kutzelnigg1979generalized} for completeness.   This approach starts from the Lie-algebraic perspective on the variational principle.  
The generators for variations in a norm conserved wavefunction are elements of a complex Lie algebra.  The variational principle which states
\begin{align}
\delta \tilde{E} = \delta \langle \tilde{\psi} \vert H \vert \tilde{\psi} \rangle = 0
\end{align}
can be cast as stationarity with respect to a unitary group
\begin{align}
U = e^{R} \;\; R = -R^{\dagger}
\end{align}
where $R$ is an element of the Lie algebra $\mathcal{L}$ supporting $H$. Formulation of the variations in $\tilde{E}$ with respect to $R$ is formulated using the BCH expansion
\begin{align}
\tilde{E} \rightarrow \tilde{E}' = \tilde{E} + \langle \tilde{\psi} \vert \left[ H, R \right] \vert \tilde{\psi} \rangle + \frac{1}{2}\langle \tilde{\psi} \vert \left[ \left[ H, R \right], R \right] \vert \tilde{\psi} \rangle + ...
\end{align}
and thus stationarity with respect to infinitesimal variations in $R$ implies
\begin{align}\label{eq:brillouin}
\langle \tilde{\psi} \vert \left[ H, R \right] \vert \tilde{\psi} \rangle = 0 \;\; \forall R = -R^{\dagger}
\end{align}
\subsection{Iteratively constructing wavefunctions}
Given an $R$ that does not satisfy the first order stationarity condition \eq{brillouin} we can propose a new wavefunction that is approximately stationary with respect to $R$.
\begin{align}
A_{R} = \langle \phi \vert \left[ H, R \right] \vert \phi \rangle \neq 0
\end{align}
We want to determine an update of the generator $R$ such that the first order condition holds.  We consider the update to the wavefunction
\begin{align}\label{eq:psi_update}
\psi = e^{-f_{R}R}\phi
\end{align}
where $f_{r}$ is a real number.  Considering how the energy changes as a function of $f_{R}$
\begin{align}
E_{f_{R}} = \langle \phi \vert e^{f_{R}R} H e^{-f_{R}R} \vert \phi \rangle \approx \langle \phi \vert H + f_{R}\left[H, R\right] + \frac{f_{R}^{2}}{2}\left[\left[H, R\right], R\right] \vert \phi\rangle
\end{align}
In a similar fashion to deriving a Newton-Raphson update in optimization we can differentiate to find an $f_{R}$ that approximately satisfies \eq{brillouin}.
\begin{align}
\frac{d E_{f_{R}}}{d f_{R}} = \langle \phi \vert \left[H, R\right] + \frac{f_{R}}{2}\left[\left[H, R\right], R \right] \vert \phi \rangle = 0 \qquad \quad
f_{R} = -A_{R} / B_{R,R} \qquad \quad
B_{R, R} = \langle \phi \vert \left[\left[H, R\right], R \right] \vert \phi \rangle
\end{align}
Alternatively, one can determine the change in the stationary condition with respect to $f_{R}$
\begin{align}
0 = \langle \phi \vert e^{f_{R}R} \left[H, R\right] e^{-f_{R}R} \vert \phi \rangle  = \langle \phi \vert \left[H, R\right] + f_{R}\left[\left[H, R\right], R\right] + \frac{f_{R}^{2}}{2}\left[\left[\left[H, R\right], R\right], R\right] +  ... \vert \phi \rangle
\end{align}
and enforce the stationarity approximately by truncating at first order and solving for $f_{R}$
\begin{align}
f_{R} = -A_{R} / B_{R, R}
\end{align}
which provides the same type of update. The error in the residual for $R$, $A_{R}$, is now of the magnitude $O(f_{R}^{2})$ at leading order.  This update inspires a possible iterative procedure for improving the wavefunction that will quadratically converge to the correct state if we are in a convex region away from the exact solution~\cite{kutzelnigg1979generalized}.

One can use the above procedure where $R$ is not an element of the operator basis $\{X_{k}\}$ of the Lie algebra $\mathcal{L}$
\begin{align}
R = \sum_{k}c_{k}X \;\;,\;\; X \subset \mathcal{L}
\end{align}
and to determine a set of $c_{k}$ which approximately satisfy \eq{brillouin}.
\begin{align}\label{eq:non_basis_update}
0 \approx \langle \phi \vert \left[H, X_{k} \right]  + \sum_{l}\left[ \left[ H, X_{k}\right], X_{l} \right]c_{l} \vert \phi \rangle
\end{align}
Again, approximating the expansion in \eq{non_basis_update} to first order we get a system of equations to solve for $c_{k}$ that ensures the Brillouin condition is satisfied up to leading error of $\mathcal{O}(c_{k}^{2})$.

In the context of a NISQ machine one needs to consider the family of generators $\{R\}$ that is tractable and the cost of the measurements associated with measuring $A_{R}$ and $B_{k, l}$.  In this work we use
\begin{align}
A_{p,q} = \langle \psi \vert \left[ H, a_{p}^{\dagger}a_{q} \right] \vert \psi \rangle \qquad \qquad
B_{p,q;r,s} = \langle \psi \vert \left[ \left[H, a_{p}^{\dagger}a_{q}\right], a_{r}^{\dagger}a_{s}\right] \vert \psi \rangle.
\end{align}
Both the gradient and the Hessian term can be evaluated with knowledge of the $1$-RDM under the assumption that $\psi$ corresponds to a Slater determinant.  The update parameters to $\kappa$, $f_{p,q}$ are computed by solving the augmented Hessian eigenvalue problem
\begin{align}
\begin{pmatrix}
0 & \mathbf{A} \\
\mathbf{A}^{\dagger} & \mathbf{B}
\end{pmatrix}
\begin{pmatrix}
1 \\
f_{p,q}
\end{pmatrix}
= \epsilon 
\begin{pmatrix}
1 \\
f_{p,q}
\end{pmatrix}
\end{align}
which provides an optimal level shift to Newton's method
\begin{align}
\mathbf{A} + (\mathbf{B} - \epsilon)f_{p,q} = 0.
\end{align}
As described in~\cite{sun2016co} we add regularization by limiting the size of the update $f_{p,q}$ by rescaling under the condition that the max update is above a parameter $\gamma$
\begin{align}
\begin{cases}
f_{p,q} & \max(f_{p,q}) < \gamma \\
\frac{\gamma}{\max(f_{p,q})}f_{p,q} & \max(f_{p,q}) \geq \gamma
\end{cases}
\end{align}
The algorithm then dictates that the wavefunction is updated through \eq{psi_update} which is yet another \nick{non-interacting} fermion wavefunction.  We concatenate this basis rotation with the original using \eq{one_body_homomorphism} so the circuit depth remains constant.  The optimization procedure is iterated for a fixed number of steps or the commutator $\langle \left[H, X_{k}\right]\rangle$ falls below a predefined threshold.

\section{Additional Performance Analysis}\label{app:syc_performance}
\subsection{Post-selection performance}
In this section we examine the percentage of measurements rejected by post-selection as a function of system size and fidelity metrics across the systems studied in the hydrogen chain and diazene experiments. In \tab{post_selection_ratios} we plot the ratio of the total number of circuit repetitions that result in the correct excitation number.  As expected this ratio decreases with system size, almost perfectly tracking a joint readout fidelity of 95\%.  We believe the \nick{discrepancy} between the two 10-qubit experiments ($\rm{H}_{10}$ and diazene experiments) stems from the fact that the diazene circuits have more idle circuit moments where the qubits are free to decay.
\begin{table}[H]
    \centering
    \begin{tabular}{|c|c|}
    \hline
    Molecule & Post-selection Shot Ratio \\
    \hline
     $\rm{H}_{6}$    & 0.764(7) \\
     $\rm{H}_{8}$    & 0.66(1) \\
     $\rm{H}_{10}$   & 0.56(1) \\
     $\rm{H}_{12}$   & 0.46(2)\\
     diazene & 0.44(1) \\
     \hline
    \end{tabular}
    \caption{The average fraction of the 250,000 circuit repetitions used to measure observables for each circuit.  The average is collected across all hydrogen geometries and diazene geometries for every circuit required to estimate the $1$-RDM for these systems.}
    \label{tab:post_selection_ratios}
\end{table}

\begin{figure}[H]
    \centering
    \includegraphics[width=8.5cm]{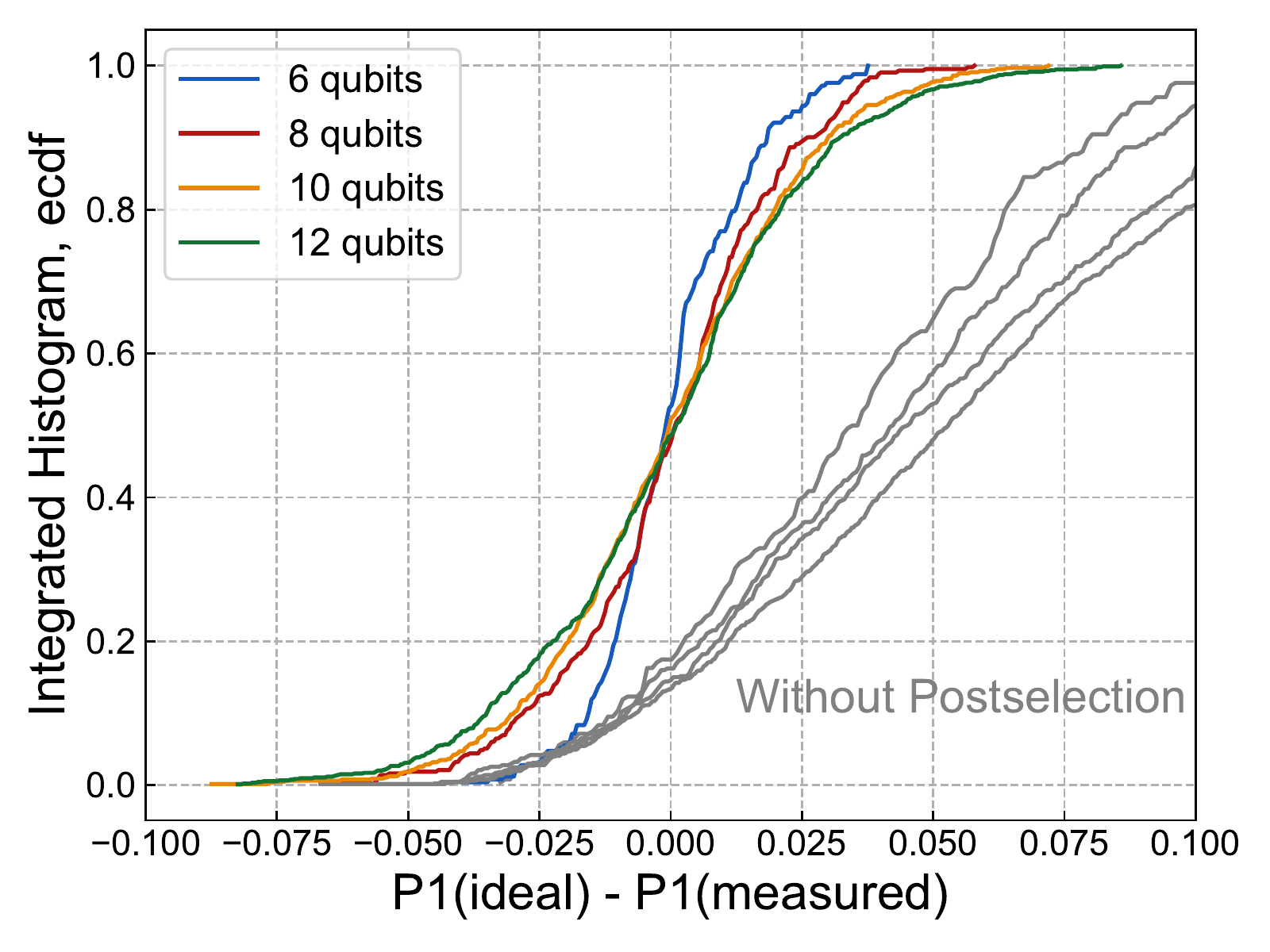}
    \caption{Integrated histogram of readout performance with and without post-selection on photon number.  Grey lines are the histograms of circuit measurements without post-selection.}
    \label{fig:static_chains_integrated_hists}
\end{figure}

Plotted another way, we can examine the distribution of local qubit expectation values $\langle M_{i} \rangle$ where $M_{i}$ is the measurement result of qubit $i$. In \figs{static_chains_integrated_hists} we plot the integrated histogram of $M_{i}$--i.e. the probability of a $1$ bit being measured from qubit $i$--(denoted P1) on all the qubits for all circuits in all hydrogen chain experiments.  This is compared to the theoretical value obtained by the perfect $1$-RDM simulation described in \app{classical_sim_of_free_fermions}.  The significant improvement in readout scatter from post-selection is a fundamental driver in the success of this experiment due to the sensitivity of quantum chemistry energies to electron number.  

\subsection{Natural Occupation Numbers}
\nick{
In this section we tabulate the natural occupation numbers for the ``raw'' and the ``post-selection'' data sets.  
\begin{table}[H]
    \centering\nick{
    \caption{H$_{6}$ raw natural orbitals}
    \begin{tabular}{|c|c|c|c|c|c|c|}
\hline
bond distance & $\lambda_1$ & $\lambda_2$ & $\lambda_3$ & $\lambda_4$ & $\lambda_5$ & $\lambda_6$\\ 
\hline
0.5 &  0.0024 &  0.0181 &  0.0338 &  0.8738 &  0.9109 &  0.9402 \\ 
0.9 &  0.0000 &  0.0176 &  0.0295 &  0.8888 &  0.9142 &  0.9405 \\ 
1.3 &  0.0083 &  0.0182 &  0.0321 &  0.8808 &  0.9157 &  0.9417 \\ 
1.7 &  0.0080 &  0.0209 &  0.0400 &  0.8780 &  0.8999 &  0.9518 \\ 
2.1 &  0.0103 &  0.0130 &  0.0378 &  0.8884 &  0.9074 &  0.9438 \\ 
2.5 &  0.0098 &  0.0128 &  0.0403 &  0.8868 &  0.9126 &  0.9427 \\ 
\hline
\end{tabular}\label{tab:h6_raw_natural_occs}}
\end{table}}

\begin{table}[H]
    \centering\nick{
    \caption{H$_{6}$ +post-selection natural orbitals}
    \begin{tabular}{|c|c|c|c|c|c|c|}
\hline
bond distance & $\lambda_1$ & $\lambda_2$ & $\lambda_3$ & $\lambda_4$ & $\lambda_5$ & $\lambda_6$\\ 
\hline
0.5 &  -0.0006 &  0.0113 &  0.0271 &  0.9768 &  0.9861 &  0.9993 \\ 
0.9 &  -0.0037 &  0.0146 &  0.0204 &  0.9815 &  0.9883 &  0.9989 \\ 
1.3 &  0.0042 &  0.0139 &  0.0266 &  0.9726 &  0.9818 &  1.0009 \\ 
1.7 &  0.0019 &  0.0164 &  0.0322 &  0.9656 &  0.9732 &  1.0106 \\ 
2.1 &  0.0075 &  0.0124 &  0.0332 &  0.9711 &  0.9737 &  1.0021 \\ 
2.5 &  0.0085 &  0.0094 &  0.0343 &  0.9689 &  0.9801 &  0.9988 \\ 
\hline
\end{tabular}\label{tab:h6_ps_natural_occs}}
\end{table}

\begin{table}[H]
    \centering\nick{
    \caption{H$_{8}$ raw natural orbitals}
    \begin{tabular}{|c|c|c|c|c|c|c|c|c|}
\hline
bond distance & $\lambda_1$ & $\lambda_2$ & $\lambda_3$ & $\lambda_4$ & $\lambda_5$ & $\lambda_6$ & $\lambda_7$ & $\lambda_8$\\ 
\hline
0.5 &  -0.0033 &  0.0083 &  0.0197 &  0.0365 &  0.8909 &  0.8975 &  0.9046 &  0.9283 \\ 
0.9 &  -0.0047 &  0.0126 &  0.0290 &  0.0435 &  0.8742 &  0.8800 &  0.9181 &  0.9331 \\ 
1.3 &  -0.0060 &  0.0148 &  0.0275 &  0.0469 &  0.8622 &  0.8881 &  0.9129 &  0.9238 \\ 
1.7 &  -0.0089 &  0.0210 &  0.0426 &  0.0538 &  0.8594 &  0.8736 &  0.9118 &  0.9179 \\ 
2.1 &  0.0026 &  0.0247 &  0.0458 &  0.0532 &  0.8454 &  0.8634 &  0.9139 &  0.9295 \\ 
2.5 &  0.0085 &  0.0259 &  0.0572 &  0.0610 &  0.8379 &  0.8801 &  0.9010 &  0.9168 \\ 
\hline
\end{tabular}\label{tab:h8_raw_natural_occs}}
\end{table}

\begin{table}[H]
    \centering\nick{
    \caption{H$_{8}$ +post-selection natural orbitals}
    \begin{tabular}{|c|c|c|c|c|c|c|c|c|}
\hline
bond distance & $\lambda_1$ & $\lambda_2$ & $\lambda_3$ & $\lambda_4$ & $\lambda_5$ & $\lambda_6$ & $\lambda_7$ & $\lambda_8$\\ 
\hline
0.5 &  -0.0087 &  0.0042 &  0.0146 &  0.0216 &  0.9779 &  0.9889 &  0.9944 &  1.0072 \\ 
0.9 &  -0.0136 &  0.0080 &  0.0207 &  0.0350 &  0.9637 &  0.9731 &  0.9992 &  1.0138 \\ 
1.3 &  -0.0160 &  0.0115 &  0.0237 &  0.0395 &  0.9535 &  0.9748 &  0.9998 &  1.0132 \\ 
1.7 &  -0.0209 &  0.0181 &  0.0357 &  0.0512 &  0.9500 &  0.9634 &  0.9926 &  1.0099 \\ 
2.1 &  -0.0120 &  0.0188 &  0.0411 &  0.0517 &  0.9458 &  0.9502 &  0.9947 &  1.0097 \\ 
2.5 &  -0.0075 &  0.0115 &  0.0513 &  0.0568 &  0.9362 &  0.9660 &  0.9814 &  1.0043 \\ 
\hline
\end{tabular}\label{tab:h8_ps_natural_occs}}
\end{table}

\begin{table}[H]
    \centering\nick{
    \caption{H$_{10}$ raw natural orbitals}
    \begin{tabular}{|c|c|c|c|c|c|c|c|c|c|c|}
\hline
bond distance & $\lambda_1$ & $\lambda_2$ & $\lambda_3$ & $\lambda_4$ & $\lambda_5$ & $\lambda_6$ & $\lambda_7$ & $\lambda_8$ & $\lambda_9$ & $\lambda_{10}$ \\
\hline
0.5 &  -0.0503 &  -0.0067 &  0.0087 &  0.0493 &  0.0638 &  0.8663 &  0.8747 &  0.8961 &  0.9109 &  0.9222 \\ 
0.9 &  -0.0176 &  0.0109 &  0.0169 &  0.0514 &  0.0829 &  0.8359 &  0.8490 &  0.8779 &  0.9165 &  0.9182 \\ 
1.3 &  -0.0050 &  0.0020 &  0.0215 &  0.0279 &  0.0460 &  0.8302 &  0.8868 &  0.8892 &  0.9142 &  0.9266 \\ 
1.7 &  -0.0289 &  0.0107 &  0.0212 &  0.0343 &  0.0535 &  0.8529 &  0.8747 &  0.8795 &  0.9106 &  0.9372 \\ 
2.1 &  -0.0092 &  0.0048 &  0.0145 &  0.0299 &  0.0596 &  0.8537 &  0.8651 &  0.8985 &  0.9217 &  0.9352 \\ 
2.5 &  -0.0010 &  0.0118 &  0.0216 &  0.0300 &  0.0626 &  0.8470 &  0.8739 &  0.8842 &  0.9107 &  0.9258 \\ 
\hline
\end{tabular}\label{tab:h10_raw_natural_occs}}
\end{table}

\begin{table}[H]
    \centering\nick{
    \caption{H$_{10}$ +post-selection natural orbitals}
    \begin{tabular}{|c|c|c|c|c|c|c|c|c|c|c|}
\hline
bond distance & $\lambda_1$ & $\lambda_2$ & $\lambda_3$ & $\lambda_4$ & $\lambda_5$ & $\lambda_6$ & $\lambda_7$ & $\lambda_8$ & $\lambda_9$ & $\lambda_{10}$ \\
\hline
0.5 &  -0.0624 &  -0.0122 &  0.0021 &  0.0492 &  0.0553 &  0.9488 &  0.9826 &  0.9984 &  1.0130 &  1.0251 \\ 
0.9 &  -0.0224 &  0.0056 &  0.0139 &  0.0408 &  0.0736 &  0.9306 &  0.9585 &  0.9830 &  0.9985 &  1.0179 \\ 
1.3 &  -0.0126 &  -0.0110 &  0.0185 &  0.0276 &  0.0497 &  0.9262 &  0.9791 &  0.9967 &  0.9978 &  1.0278 \\ 
1.7 &  -0.0397 &  -0.0005 &  0.0170 &  0.0290 &  0.0561 &  0.9470 &  0.9731 &  0.9825 &  1.0077 &  1.0276 \\ 
2.1 &  -0.0215 &  0.0026 &  0.0057 &  0.0224 &  0.0529 &  0.9559 &  0.9583 &  0.9851 &  1.0069 &  1.0317 \\ 
2.5 &  -0.0184 &  0.0057 &  0.0232 &  0.0248 &  0.0597 &  0.9441 &  0.9672 &  0.9817 &  0.9997 &  1.0122 \\ 
\hline
\end{tabular}\label{tab:h10_ps_natural_occs}}
\end{table}

\begin{table}[H]
    \centering\nick{
    \caption{H$_{12}$ raw natural orbitals}
    \begin{tabular}{|c|c|c|c|c|c|c|c|c|c|c|c|c|}
\hline
bond distance & $\lambda_1$ & $\lambda_2$ & $\lambda_3$ & $\lambda_4$ & $\lambda_5$ & $\lambda_6$ & $\lambda_7$ & $\lambda_8$ & $\lambda_9$ & $\lambda_{10}$  & $\lambda_{11}$ & $\lambda_{12}$  \\
\hline
0.5 &  -0.0037 &  0.0010 &  0.0077 &  0.0182 &  0.0500 &  0.0589 &  0.8361 &  0.8515 &  0.8828 &  0.8910 &  0.8995 &  0.9098 \\ 
0.9 &  -0.0195 &  0.0021 &  0.0118 &  0.0247 &  0.0413 &  0.0713 &  0.7948 &  0.8316 &  0.8816 &  0.8919 &  0.9019 &  0.9442 \\ 
1.3 &  -0.0160 &  0.0066 &  0.0192 &  0.0346 &  0.0521 &  0.0823 &  0.8035 &  0.8179 &  0.8853 &  0.8911 &  0.9099 &  0.9198 \\ 
1.7 &  -0.0016 &  0.0087 &  0.0276 &  0.0288 &  0.0458 &  0.0737 &  0.7967 &  0.8480 &  0.8614 &  0.8906 &  0.8991 &  0.9099 \\ 
2.1 &  -0.0153 &  -0.0011 &  0.0235 &  0.0325 &  0.0572 &  0.0777 &  0.8198 &  0.8331 &  0.8556 &  0.8646 &  0.9132 &  0.9260 \\ 
2.5 &  -0.0143 &  0.0029 &  0.0207 &  0.0564 &  0.0650 &  0.0821 &  0.8143 &  0.8351 &  0.8719 &  0.8750 &  0.8966 &  0.9131 \\ 
\hline
\end{tabular}\label{tab:h12_raw_natural_occs}}
\end{table}

\begin{table}[H]
    \centering\nick{
    \caption{H$_{12}$ +post-selection natural orbitals}
    \begin{tabular}{|c|c|c|c|c|c|c|c|c|c|c|c|c|}
\hline
bond distance & $\lambda_1$ & $\lambda_2$ & $\lambda_3$ & $\lambda_4$ & $\lambda_5$ & $\lambda_6$ & $\lambda_7$ & $\lambda_8$ & $\lambda_9$ & $\lambda_{10}$  & $\lambda_{11}$ & $\lambda_{12}$  \\
\hline
0.5 &  -0.0120 &  -0.0029 &  0.0030 &  0.0137 &  0.0474 &  0.0536 &  0.9349 &  0.9540 &  0.9878 &  0.9968 &  1.0042 &  1.0194 \\ 
0.9 &  -0.0310 &  -0.0039 &  0.0037 &  0.0222 &  0.0337 &  0.0735 &  0.8947 &  0.9397 &  0.9889 &  0.9970 &  1.0166 &  1.0649 \\ 
1.3 &  -0.0260 &  -0.0009 &  0.0113 &  0.0216 &  0.0454 &  0.0817 &  0.8979 &  0.9265 &  0.9924 &  0.9976 &  1.0231 &  1.0292 \\ 
1.7 &  -0.0074 &  0.0002 &  0.0147 &  0.0231 &  0.0459 &  0.0719 &  0.8964 &  0.9596 &  0.9665 &  0.9995 &  1.0082 &  1.0215 \\ 
2.1 &  -0.0296 &  -0.0082 &  0.0172 &  0.0223 &  0.0593 &  0.0735 &  0.9170 &  0.9403 &  0.9715 &  0.9803 &  1.0134 &  1.0430 \\ 
2.5 &  -0.0252 &  -0.0156 &  0.0096 &  0.0397 &  0.0618 &  0.0878 &  0.9139 &  0.9410 &  0.9782 &  0.9837 &  1.0029 &  1.0222 \\ 
\hline
\end{tabular}\label{tab:h12_ps_natural_occs}}
\end{table}

\subsection{Energy and Fidelity}

\begin{figure}[H]
    \centering
    \includegraphics[width=8.5cm]{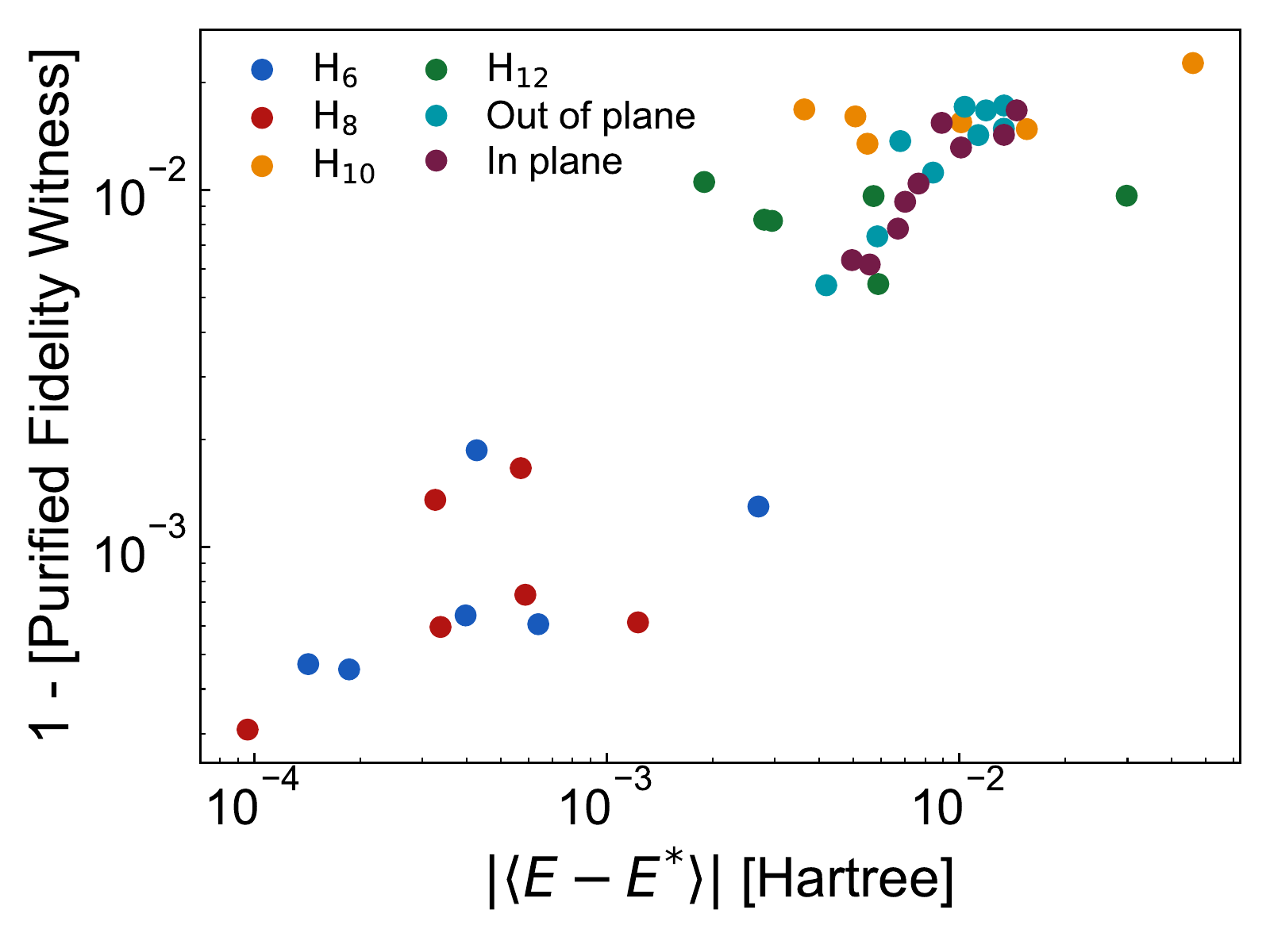}
    \caption{Absolute error versus fidelity witness for VQE optimized with error mitigation for all experiments.}
    \label{fig:mae_fidelity}
\end{figure}

In \figs{mae_fidelity} we plot the log-log scatter of absolute error and fidelity witness for all systems studied. The correlation in the fidelity and absolute energy error suggests that fidelity can be used as an optimization target for this system.  This is a useful property when considering basis rotation states as targets for benchmarks and tune-up protocols.  
Slow two-level-system (TLS) diffusion on the surface of superconducting processors can alter the performance of qubits over time periods of hours or days. It is likely that the H$_{10}$ data set was collected during a time where the best performing qubits had worse coupling to an itinerant TLS\cite{PhysRevLett.121.090502} than when we collected the H$_{12}$ dataset. Thus, there was some variance in performance across the different days when the chip was used to collect data. We believe that by showing all of these results without cherry picking and rerunning less performant curves, we give a more accurate representation of the average performance of the device.

To better describe the consistent quality of VQE optimized 10 qubit calculations we tabulate the perceived fidelity calculated from purified $1$-RDMs in all 10 qubit experiments: six H$_{10}$ experiments and eighteen diazene points.  On all but one experiment variational relaxation combined with other error mitigation techniques allows us to achieve $>98.0\%$ average fidelity.
\begin{figure}[H]
    \centering
    \includegraphics[width=8.5cm]{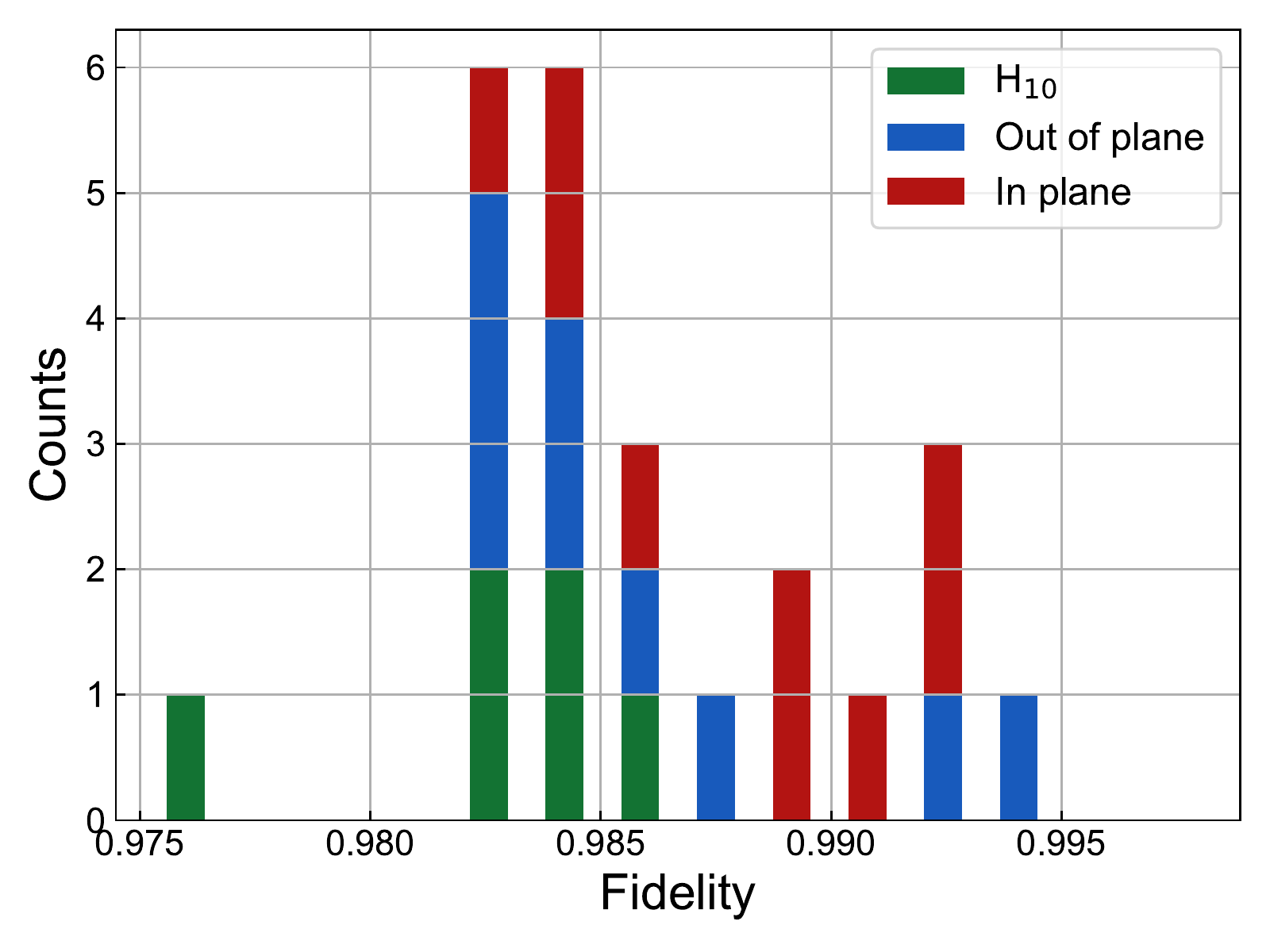}
    \caption{\textbf{\textit{Fidelity of 10 qubit experiments:}} A histogram of fidelity witness values associated with the VQE optimized 10 qubit systems. }
    \label{fig:ten_q_fidelities}
\end{figure}
\section{Molecular geometries}
For the hydrogen chains OpenFermion~\cite{openfermion} and Psi4~\cite{Psi4} were used to generate the integrals.  All hydrogen chains were computed at atom-atom separations of 0.5, 0.9, 1.3, 1.7, 2.1, and 2.5 $\textrm{\AA}$. For the diazene curves we used Psi4 to map out the reaction coordinate for each isomerization mechanism by optimizing the geometries of the molecule \nick{simultaneously} constraining either the dihedral angle or NNH angle to a fixed value.  \tab{outofplane} and \tab{inplane}, below, contain the geometries we considered for out-of-plane rotation and in-plane rotation of the hydrogen atom.   To reduce diazene to a 10 qubit problem we perform two cycles of canonical Hartree-Fock self-consistent field and then integrate out the bottom two energy levels. 
\begin{center}
\begin{table}[h]
\begin{minipage}[b]{0.4\textwidth}
\caption{Out-of-plane rotation geometries \label{tab:outofplane}}
\begin{tabular}{|c|c|c|c|c|}
\hline
Internal coord. & Atom & \multicolumn{3}{c|}{Cartesian coordinates} \\
\hline
3.157 & H & -0.00183 & 0.61231 & -1.23326 \\ 
  & N & -0.00183 & 0.61231 & -0.16961 \\ 
  & N & -0.00183 & -0.56366 & 0.29317 \\ 
  & H & 0.05269 & -1.28820 & -0.48362 \\ 
 & & & & \\
26.315 & H & -0.01473 & 0.61213 & -1.23797 \\ 
  & N & -0.01473 & 0.61213 & -0.17381 \\ 
  & N & -0.01473 & -0.56586 & 0.29104 \\ 
  & H & 0.42406 & -1.25504 & -0.39080 \\ 
 & & & & \\
49.473 & H & -0.02522 & 0.61175 & -1.25087 \\ 
  & N & -0.02522 & 0.61175 & -0.18596 \\ 
  & N & -0.02522 & -0.57104 & 0.28761 \\ 
  & H & 0.72616 & -1.17742 & -0.16152 \\ 
 & & & & \\
72.631 & H & -0.03098 & 0.60150 & -1.29530 \\ 
  & N & -0.03098 & 0.60150 & -0.23059 \\ 
  & N & -0.03098 & -0.56623 & 0.30717 \\ 
  & H & 0.89199 & -1.09153 & 0.23125 \\ 
 & & & & \\
95.641 & H & -0.03338 & 0.62184 & -1.24521 \\ 
  & N & -0.03338 & 0.62184 & -0.18592 \\ 
  & N & -0.03338 & -0.60178 & 0.24302 \\ 
  & H & 0.96106 & -0.90055 & 0.45184 \\ 
 & & & & \\
117.611 & H & -0.03017 & 0.62034 & -1.22808 \\ 
  & N & -0.03017 & 0.62034 & -0.16858 \\ 
  & N & -0.03017 & -0.60975 & 0.20354 \\ 
  & H & 0.86843 & -0.76736 & 0.74227 \\ 
 & & & & \\
139.581 & H & -0.02217 & 0.61500 & -1.22656 \\ 
  & N & -0.02217 & 0.61500 & -0.16715 \\ 
  & N & -0.02217 & -0.61052 & 0.18284 \\ 
  & H & 0.63829 & -0.67733 & 1.00847 \\ 
 & & & & \\
161.551 & H & -0.01085 & 0.61243 & -1.22466 \\ 
  & N & -0.01085 & 0.61243 & -0.16601 \\ 
  & N & -0.01085 & -0.61150 & 0.16941 \\ 
  & H & 0.31225 & -0.62540 & 1.17746 \\ 
 & & & & \\
183.522 & H & 0.00211 & 0.61159 & -1.22471 \\ 
  & N & 0.00211 & 0.61159 & -0.16627 \\ 
  & N & 0.00211 & -0.61155 & 0.16640 \\ 
  & H & -0.06064 & -0.61209 & 1.22297 \\
\hline
\end{tabular}
\end{minipage}
\begin{minipage}[b]{0.4\textwidth}
\caption{In-plane rotation geometries \label{tab:inplane}}
\begin{tabular}{|c|c|c|c|c|}
\hline
Internal Coord. & Atom & \multicolumn{3}{c|}{Cartesian coordinates} \\
\hline
108.736 & H & 0.00000 & 0.61228 & -1.23237 \\ 
  & N & 0.00000 & 0.61228 & -0.16925 \\ 
  & N & 0.00000 & -0.56613 & 0.29515 \\ 
  & H & 0.00001 & -1.25344 & -0.51686 \\ 
 & & & & \\
127.473 & H & 0.00000 & 0.61339 & -1.24223 \\ 
  & N & 0.00000 & 0.61339 & -0.17528 \\ 
  & N & 0.00000 & -0.55235 & 0.28364 \\ 
  & H & 0.00001 & -1.46143 & -0.26340 \\ 
 & & & & \\
146.210 & H & 0.00000 & 0.61423 & -1.26614 \\ 
  & N & 0.00000 & 0.61423 & -0.18644 \\ 
  & N & 0.00000 & -0.54592 & 0.27365 \\ 
  & H & 0.00001 & -1.56334 & 0.05447 \\ 
 & & & & \\
164.947 & H & 0.00000 & 0.61854 & -1.26711 \\ 
  & N & 0.00000 & 0.61854 & -0.18132 \\ 
  & N & 0.00000 & -0.55047 & 0.24812 \\ 
  & H & 0.00000 & -1.56434 & 0.33892 \\ 
 & & & & \\
182.0 & H & 0.00000 & 0.62468 & -1.24838 \\ 
  & N & 0.00000 & 0.62468 & -0.16646 \\ 
  & N & 0.00000 & -0.56138 & 0.21599 \\ 
  & H & 0.00007 & -1.50420 & 0.56008 \\ 
 & & & & \\
200.526 & H & -0.00002 & 0.63051 & -1.22163 \\ 
  & N & -0.00002 & 0.63051 & -0.14939 \\ 
  & N & -0.00002 & -0.57522 & 0.18140 \\ 
  & H & 0.00060 & -1.39873 & 0.77683 \\ 
 & & & & \\
219.052 & H & -0.00004 & 0.63081 & -1.20416 \\ 
  & N & -0.00004 & 0.63081 & -0.14048 \\ 
  & N & -0.00004 & -0.58876 & 0.15662 \\ 
  & H & 0.00116 & -1.21501 & 0.97997 \\ 
 & & & & \\
237.578 & H & -0.00005 & 0.62175 & -1.20948 \\ 
  & N & -0.00005 & 0.62175 & -0.14932 \\ 
  & N & -0.00005 & -0.59932 & 0.15347 \\ 
  & H & 0.00155 & -0.93328 & 1.15189 \\ 
 & & & & \\
256.105 & H & -0.00004 & 0.61032 & -1.22743 \\ 
  & N & -0.00004 & 0.61032 & -0.16648 \\ 
  & N & -0.00004 & -0.61200 & 0.16650 \\ 
  & H & 0.00111 & -0.58711 & 1.22715 \\ 
 \hline
\end{tabular}
\end{minipage}
\end{table}
\end{center}

\end{document}